\documentclass[reqno,11pt]{amsart}
\usepackage{hyperref}
\usepackage[mathscr]{eucal}
\usepackage{enumerate}
\numberwithin{equation}{section}

\newtheorem{Def}{Definition}[section]
\newtheorem{Thm}[Def]{Theorem}

\newtheorem{Lemma}[Def]{Lemma}

\newcommand{\beq}{\begin{equation}}
\newcommand{\eeq}{\end{equation}}
\newcommand{\Proof}{\begin{proof}}
\newcommand{\QED}{\end{proof} \noindent}

\newcommand{\R}{\mathbb{R}}

\allowdisplaybreaks[1]

\title[The Scattering of Gravity Waves by regularity singularities]
{``Regularity Singularities'' \\ and \\ the Scattering of Gravity Waves in Approximate Locally Inertial Frames}

\author[M.\ Reintjes]{Moritz Reintjes}
\address{IMPA - Instituto Nacional de Matem{\'a}tica Pura e Aplicada \\ Rio de Janeiro, 22460-320, Brasil}
\email{moritzreintjes@gmail.com}
\thanks{M. Reintjes was supported by the Deutsche Forschungsgemeinschaft (DFG), Grant Number RE 3471/2-1, from January 2013 until December 2014. Since January 2015 M. Reintjes is a Post-Doctorate at IMPA, funded through CAPES-Brazil.}

\author[B.\ Temple]{Blake Temple \\ \\ June 9, 2016}
\address{Department of Mathematics\\ University of California\\ Davis, CA 95616\\ USA}
 \email{temple@math.ucdavis.edu}
\thanks{B. Temple was supported by NSF Applied Mathematics Grant Number DMS-010-2493.}

\begin{document}

\maketitle

\begin{abstract}
It is an open question whether solutions of the Einstein-Euler equations are smooth enough to admit locally inertial coordinates at points of shock wave interaction, or whether  ``regularity singularities'' can exist at such points.  The term {\it regularity singularity} was proposed by the authors as a point in spacetime where the gravitational metric tensor is Lipschitz continuous ($C^{0,1}$), but no smoother, in any coordinate system of the $C^{1,1}$ atlas.   An existence theory for shock wave solutions in $C^{0,1}$ admitting arbitrary interactions has been proven for the Einstein-Euler equations in spherically symmetric spacetimes, but $C^{1,1}$ is the requisite smoothness required for space-time to be locally flat. Thus the open problem of regularity singularities is the problem as to whether locally inertial coordinate systems exist at shock waves within the larger $C^{1,1}$ atlas.   To clarify this open problem, we identify new ``Coriolis type'' effects in the geometry of $C^{0,1}$ shock wave metrics and prove they are essential in the sense that they can never be made to vanish within the atlas of {\it smooth} coordinate transformations, the atlas usually assumed in classical differential geometry.  Thus the problem of existence of regularity singularities is equivalent to the question as to whether or not these Coriolis type effects are essentially non-removable and `real', or merely coordinate effects that can be removed, (in analogy to classical Coriolis forces), by going to the less regular atlas of $C^{1,1}$ transformations.  If essentially non-removable, it would argue strongly for a `real' new physical effect for General Relativity, providing a physical context to the open problem of regularity singularities.
\end{abstract}


\section{Introduction}\label{intro}

A basic question for the shock wave theory in General Relativity (GR) is the regularity of the gravitational metric at shock waves.   Shock waves are solutions of the Einstein-Euler equations in which the fluid density and velocity are discontinuous.  Shock waves always form in solutions of the compressible Euler equations whenever the flow is sufficiently compressive, \cite{Lax,Smoller}.  Shock waves introduce increase of entropy, time-irreversibility and loss of information into GR, and they also create discontinuities in the curvature tensors of space-time.   Classical shock waves in non-relativistic gas dynamics are regularized by shock profiles when viscosity and heat conduction are included,  but the theory of dissipation is problematic in relativity due to the fact that parabolic equations introduce infinite speed of propagation, and modified theories of dissipation have been controversial as they typically are either not causal, or do not admit shock profiles (see \cite{FreistuehlerTemple} for references).\footnote{See \cite{FreistuehlerTemple} for a new theory of dissipation based on including relativistic viscosity and heat conduction parameters to obtain a symmetric hyperbolic regularization of the classical relativistic Navier-Stokes-Fourier equations for pure radiation, such that the resulting equations are causal and dissipative, and such that all shocks admit shock profiles.} Moreover, the discontinuities that appear in the zero dissipation limit are replaced by steep gradients near the limit, and the essential issues of shock waves persist.  Thus the Einstein-Euler equations have played a fundamental role in relativity both because they accurately model highly relativistic flows, and because, at a fundamental level, shock waves describe an accurate idealized limit which introduces dissipation into relativity without giving up causality.  Even so, it remains an open problem as to the regularity of the gravitational metric at GR shock waves.    Specifically, it is not known whether the space-time is smooth enough to admit locally inertial coordinates in which the metric is Minkowski and the derivatives of the metric vanish at points of shock wave {\it interaction}.\footnote{We say a point of shock wave interaction is a point where multiple waves interact in any complicated fashion, so long as one of them is a shock wave.}   If space-time is not locally inertial in the zero dissipation limit, then the lack of existence of locally inertial frames in this limit would persist as an issue under perturbation by shock profiles because the jumps in first-order derivatives would get replaced by large second-order derivatives, and the irregularity of the metric would simply be spread out over a small region which would appear to be a regularity singularity in the far field limit.

A basic existence theory for shock wave interactions in the Einstein-Euler equations based on Glimm's method was established in \cite{GroahTemple} for spherically symmetric spacetimes,  and interestingly, the methods are only sufficient to prove existence for gravitational metrics $g$ which are only Lipschitz continuous. It is not known whether the metrics associated with solutions in \cite{GroahTemple}, or any more general shock wave solutions, can be smoothed from $C^{0,1}$ to $C^{1,1}$ by coordinate transformation at points of complicated shock wave interaction.\footnote{Here $C^{0,1}$ denotes continuous with Holder derivative one, (i.e., Lipschitz continuous, \cite{Evans}),  so for metrics $g\in C^{0,1}$, first derivatives of the metric suffer a jump discontinuity at shocks, while for $g\in C^{1,1}$ the second derivative suffers a jump discontinuity at shocks.}  The metric regularity $C^{1,1}$ is the minimum regularity that guarantees a metric admits locally inertial coordinate frames, and for the weak and strong formulation of the Einstein equations to be equivalent, \cite{SmollerTemple}.  Moreover, within the $C^{1,1}$ atlas, the condition that solutions be free of delta function sources is a covariant condition, and this appears to be the weakest atlas with this property, appropriate for shock wave solutions in GR,  c.f. \cite{SmollerTemple}.  

To set up a framework for addressing the question as to whether metrics associated with shock waves can be smoothed from $C^{0,1}$ to $C^{1,1}$ by coordinate transformation, we begin below by proving that if $g$ is a shock wave solution which is only $C^{0,1}$ regular in one coordinate system, then the metric cannot be smoothed to $C^{1,1}$ within the atlas of {\it smooth} (say, $C^{2,1}$) coordinate transformations, the atlas usually assumed in classical differential geometry. Therefore such metrics do not admit locally inertial coordinate frames within the smooth atlas, (c.f. Theorem \ref{Thm_app} below).  However, the Einstein equations remain consistent in the weak sense when the smooth atlas is extended to the larger atlas of $C^{1,1}$ coordinate transformations, \cite{SmollerTemple}. The Jacobians of $C^{1,1}$ transformations are only $C^{0,1}$, and thus $C^{1,1}$ transformations have the required properties to potentially eliminate jumps in the metric derivatives at shocks.  Thus the more singular atlas of $C^{1,1}$ coordinate transformations is the atlas that holds the possibility of lifting the metric regularity to $C^{1,1}$, \cite{Israel,SmollerTemple}.     Thus a most natural open question is whether the gravitation metric can be smoothed from $C^{0,1}$ to $C^{1,1}$ by $C^{1,1}$ coordinate transformations, and this addresses the locally flat character of space-time at GR shock waves. 

Following our work in \cite{ReintjesTemple}, we define a {\it regularity singularity} as a point in space-time where the metric does not admit locally inertial frames within the $C^{1,1}$ atlas, and it is an open problem as to whether regularity singularities can be created by shock wave interactions in GR.   The authors know of no physical or mathematical principle that can rule out regularity singularities in the Einstein-Euler equations ahead of time, and only mathematical proof can ultimately resolve the issue.

The starting point for our analysis here is the celebrated paper \cite{Israel} by Israel, in which he shows that at smooth co-dimension one shock surfaces in $n$ dimensions, the gravitational metric can always be smoothed from $C^{0,1}$ to $C^{1,1}$ by introducing Gaussian normal coordinates (GNC) at the shock.  The transformation to GNC is a $C^{1,1}$ transformation \cite{SmollerTemple}.  GNC are only defined for single, non-interacting shock surfaces and do not exist for the more complicated $C^{0,1}$ metrics constructed in the Groah-Temple framework \cite{GroahTemple}.   Since Israel's result in \cite{Israel}, it has been unknown whether the regularity of the gravitational metric can always be smoothed from $C^{0,1}$ to $C^{1,1}$ by coordinate transformation at shock wave interactions.   Thus it remains an outstanding open problem as to whether space-time is always locally flat at points of shock wave interaction in GR.

The first extension of Israel's theorem to the more complicated setting of shock wave {\it interactions} was accomplished by the authors in their recent papers \cite{ReintjesTemple2,Reintjes}.   The proof demonstrates in the first case ever that the gravitational metric {\it can} always be smoothed from $C^{0,1}$ to $C^{1,1}$ at a point of interacting shock waves in GR, namely the case of regular shock wave interaction in spherical symmetry between shocks from different characteristic families.  The proof introduces an explicit physical procedure for finding the coordinates that display the physics in their simplest form, replacing the GNC construction in an essential new way, based on solving a non-local hyperbolic-type system of equations. But the argument is tailored to the specific case of two interacting shock waves, and there are many miracles in the proof, places where the Rankine-Hugoniot jump conditions come in to surprisingly make an apparently over-determined system barely solvable. Thus it is not clear whether or how this proof can be extended to more complicated shock wave interactions.  For complicated shock interactions, including the more complicated $C^{0,1}$ solutions that exist by \cite{GroahTemple}, or for complicated asymmetric shock interactions in $(3+1)$-dimensions, the question as to the locally flat nature of space-time and whether regularity singularities can be created by shock wave interactions, remains an open problem.

  
Our work on the problem has led us to conjecture that if regularity singularities actually exist,  the structure of space-time at the singularity would be essentially determined by its structure within the smooth atlas alone.  That is, at a regularity singularity, the larger $C^{1,1}$ atlas would offer no essential improvement in the regularity of the gravitational metric over and above that observed in the $C^{2,1}$ atlas.
Thus to begin studying the implications of regularity singularities, should they actually exist,  we first determine properties  of a Lipschitz continuous space-time metric within the smooth $C^{2,1}$ atlas in a neighborhood of a point on a single shock surface. These properties extend easily to the case of shock wave interaction.   Our purpose in this paper, then, is to establish physical implications of the assumption that the gravitational metric is no more regular than its regularity within the smooth atlas. 

In Theorem \ref{Thm_app} we prove that, restricting to the smooth atlas for a metric Lipschitz continuous across shock surfaces, the closest one can get to a locally inertial coordinate system is one which is {\it approximate locally inertial} in a natural sense we make precise.  We then characterize what we interpret as {\it Coriolis} type effects in approximate locally inertial coordinates, effects which arise from terms which only vanish in a true locally inertial coordinate system, {\it should one exist}.  These Coriolis type effects are analogous to classical Newtonian Coriolis forces which are due to terms in the gravitational force law which arise from the rotation of the earth, but would vanish in a true locally inertial coordinate system.  (Keep in mind that classical Coriolis forces would be treated as {\it real} until inertial coordinates, which remove them, are identified.)  In Section \ref{sec: Coriolis type effects}, we identify the Coriolis terms in the geodesic equations in approximate locally inertial frames. In Theorem \ref{Prop1} of Section \ref{sec: perturbed EFE}, we derive a canonical form of the linearized Einstein equations in approximate locally inertial coordinate systems and use this to identify the Coriolis terms associated with gravity waves. In Theorem \ref{thmclk}, we use our formulation of the linearized Einstein equations to prove that these Coriolis terms are nonzero and cannot be removed by coordinate transformation to any approximate locally inertial frame.   This quantifies the contributions to the scattering of gravitational radiation in approximate locally inertial coordinates. The main results are summarized in Theorem \ref{Thmintro}. Theorem \ref{Thmintro} is interesting in its own right, because it describes how far a metric Lipschitz continuous across shock surfaces is from being locally inertial within the smooth atlas, in terms of gravitational radiation. Within this context, the open problem of regularity singularities, then, is the problem as to whether locally inertial coordinate systems and the essential regularity of space-time can be improved upon by extending the smooth atlas to the larger $C^{1,1}$ atlas. We conclude that, if no such improvement exists,  then the quantifiable effects produced by these non-removable Coriolis terms are physical implications of regularity singularities.  
\footnote{The assumption that the curvature tensors contain no delta function sources is crucial for the conjecture as to whether regularity singularities exist in GR.  Shock surfaces, by definition, are weak solutions which contain no delta function sources. Interfaces which contain delta function sources were introduced in Israel's theory of thin shells, and these include the {\it domain wall} between the false vacuum and the true vacuum in Guth's celebrated theory of inflation, \cite{guth}.  Metrics with interfaces containing delta function sources  lie in $C^{0,1}$, but cannot be smoothed to $C^{1,1}$ within the $C^{1,1}$ atlas {\it at the start} because $C^{1,1}$ metrics have classical curvature tensors which contradict the existence of delta function sources. But, when the curvature tensor is free of delta function sources, there appears to be no physical reason for this loss of regularity.}

\section{Preliminaries}

In General Relativity, the gravitational field is described by a Lorentzian metric $g$ of signature ~$(-1,1,1,1)$ on a four-dimensional spacetime manifold $M$. We call $M$ a $C^k$-manifold if it is endowed with a $C^k$-atlas, a collection of four-dimensional local diffeomorphisms that are $C^k$ regular from $M$ to $\R^4$. A composition of two local diffeomorphisms $x$ and $y$ of the form ~$x\circ y^{-1}$ is referred to as a coordinate transformation. 

 Our index notation for tensors here sometimes uses indices to determine the coordinate system, e.g., ~$T^\mu_\nu$ denotes a $(1,1)$-tensor in coordinates $x^\mu$ and ~$T^\alpha_\beta$ denotes the same tensor in coordinates $x^\alpha$. 
We use the Einstein summation convention whereby repeated up-down indices are summed over all values for the given indices. Tensors transform by contraction with the Jacobian $J^\mu_\alpha= \tfrac{\partial x^\mu}{\partial x^\alpha}$ and the inverse Jacobian, $(J^{-1})^\alpha_\nu$, which we denote by $J^\alpha_\nu$ whenever there is no confusion. In particular, the metric transforms as  $g_{\mu\nu}= J^\alpha_\mu J^\beta_\nu g_{\alpha\beta}$. Tensor-indices are raised and lowered with the metric $g_{\mu\nu}$ and its inverse  $g^{\mu\nu}$.

The time evolution of a gravitational field in general relativity is governed by the \emph{Einstein equations} \cite{Einstein} 
\beq\label{EFE} 
G^{ij}=\kappa T^{ij}, 
\eeq
a system of $10$ second-order partial differential equations that relate the metric tensor $g_{ij}$ to the undifferentiated sources $T^{ij}$ through the Einstein curvature tensor
\beq\label{Einstein tensor} 
G^{ij}=R^{ij}-\frac12 R g^{ij},
\eeq
a tensor involving second derivatives of $g$.  The Ricci tensor, $R_{jl}$, is the trace of the Riemann tensor, and the scalar curvature, $R$, is the trace of the Ricci tensor, that is, 
$$
R_{jl}= g^{ik}R_{ijkl} \ \ \ \ \ \text{and} \ \ \ R= g^{jl}R_{jl}.
$$
The Riemann tensor is given in terms of the Christoffel symbols $\Gamma^i_{lj}$ by the formula
\beq\label{Riemann_prelim} 
R_{ijkl}=  \Gamma_{ijk,l} - \Gamma_{ijl,k}+ g^{\sigma\rho} \left( \Gamma_{i \sigma l} \Gamma_{\rho jk} - \Gamma_{i \sigma k} \Gamma_{\rho jl} \right),
\eeq 
where 
\beq \label{Chr_sym_prelim} 
\Gamma_{l \,ij} = \frac12 \left( g_{li,j} + g_{jl,i} - g_{ij,l} \right) \ \ \ \ \text{and}  \ \ \ \Gamma^k_{\ ij} = g^{kl}\Gamma_{l \,ij}\, .
\eeq    
and ~$\kappa=-\frac{8\pi}{c^4} \mathcal{G}$ is the coupling constant which incorporates Newton's gravitational constant $\mathcal{G}$ and the speed of light $c$.\footnote{Note that because the curvature tensor is anti-symmetric in $(k,l)$ it requires a choice of sign, and this choice is not uniform in the literature.  Here we use Weinberg's convention \cite{Weinberg}, also used by the {\it tensor} package of MAPLE,  and with this sign convention for the curvature tensor, the gravitational constant $\kappa$ is negative, i.e.,  ~$\kappa=-\frac{8\pi}{c^4} \mathcal{G}$.  The careful reader should beware that Hawking and Ellis, as well as the MAPLE {\it geometry} package, use the opposite sign convention, defining $R$ to be minus our $R$, resulting in ~$\kappa=\frac{8\pi}{c^4} \mathcal{G}$.  Note that Weinberg's convention $\eta=(1,-1,-1,-1)$, different from ours, has no bearing on this choice of sign.  We also point out that our methods apply essentially unchanged for non-zero cosmological constant.}      Here $T^{ij}$ is the energy-momentum tensor.
For our methods we do not need to specify the matter sources, but our motivation comes from the case when $T^{ij}$ is the energy-momentum tensor for a perfect fluid,
 \beq\label{EulerStress}
 T^{ij}=(p+\rho)u_i u_j + p g_{ij},
 \eeq
 where $\rho$ is the energy density, $u_i$ the unit $4$-velocity, $u^iu_i=-1$, and $p$ the pressure.  
 
Conservation of energy and momentum enter the Einstein equations through
\beq \label{Euler}
T^{ij}_{\ \ ;j}=0\, ,
\eeq
which reduces to the relativistic compressible Euler equations in flat spacetime, and follows from the divergence-free property of the Einstein equations,
~$G^{ij}_{\ \ ;j}=0$,
a property built into $G$ at the start as an identity following from the Bianchi identities of geometry, \cite{Weinberg}.  Here as usual, semicolon denotes covariant differentiation 
$$v^i_{;j}= v^i_{,j}+ \Gamma^i_{lj}v^l,$$
where $\Gamma^i_{lj}$ denote the Christoffel symbols associated with metric $g$, defined in \eqref{Chr_sym_prelim}.
In the special case of a perfect fluid, equations \eqref{EFE} and \eqref{Euler} form the coupled Einstein-Euler equations; a system of second-order differential equations for the unknown metric $g_{ij}$, coupled to the fluid variables $\rho$, $p$ and $u^j$, a system which closes upon specification of the equation of state, \cite{Weinberg}.   

In special relativity the spacetime metric is taken to be $g_{ij}\equiv\eta_{ij}$ where $\eta_{ij}={\it diag}(-1,1,1,1)$ is the Minkowski metric, in which case (\ref{Euler}) reduces to the relativistic compressible Euler equations, a system of conservation laws in which it is well-known that shock waves form out of smooth initial data whenever the flow is sufficiently compressive. Shock waves are discontinuous solutions that solve the Euler equations weakly, in a distributional sense, \cite{Lax,Smoller}.   Across a smooth shock surface $\Sigma$, the Rankine-Hugoniot jump conditions hold,
\beq
\label{JC}
[T^{ij}] n_j =0,
\eeq
where $[f]=f_L-f_R$ denotes the jump from right to left in function $f$ across $\Sigma$, and $n_{j}$ is the surface normal.  In particular, for smooth shock surfaces, the jump conditions (\ref{JC}) are equivalent to the shock wave solution satisfying the weak formulation of (\ref{Euler}) across $\Sigma$, c.f. \cite{Smoller}.

\section{The smooth ($C^{2,1}$) atlas at points of shock wave interaction and approximate locally inertial frames} \label{sec: smooth atlas}

At a point $p$ where the space-time metric is at least $C^{1,1}$, the Riemann normal coordinate construction implies that one can always transform the metric to locally inertial coordinates at $p$, coordinates for which $g_{ij}(p)=\eta_{ij}\equiv diag(-1,1,1,1)$ and $g_{\alpha\beta,\gamma}(p) = 0$ so that the metric can be Taylor-expanded and written as
\beq\label{locally inertial metric}
g_{\alpha\beta} = \eta_{\alpha\beta} + O(\delta^2),
\eeq
where $\delta$ is the 4-dimensional (Euclidean, non-covariant) coordinate distance to $p$. That is,
\begin{eqnarray}\label{deltaq}
\delta(q)\equiv\delta(x(q)-x(p)) = \sqrt{\sum\limits_{\alpha=0}^3 \big| x^\alpha(q)-x^\alpha(p) \big|^2}.
\end{eqnarray} 

In Theorem \ref{Thm_app} below we show that the closest you can get to a locally inertial coordinate frame within the atlas of smooth (at least $C^{2,1}$) coordinate transformations for a Lipschitz metric at a point of shock wave interaction is an {\it approximate locally inertial coordinate system}, a term we make precise in the following definition.

\begin{Def} \label{defapp}
We call a coordinate system $x^\alpha$ an ``approximate locally inertial coordinate system'' at a point p if the metric takes the form 
$$g=\eta + \bar{g} $$
in a neighborhood $V$ of $p$ where $\bar{g}$ is a Lipschitz continuous symmetric tensor vanishing at $p$,  satisfying the condition that there exists an open set $U\subset V$ containing $p$ in its closure, constants $M,\bar{M}>0$, and indices $\alpha,\beta,\gamma\in\left\{0,...,3\right\}$, such that $\bar{g}$ is smooth\footnote{In all of this paper $\bar{g} \in C^{1,1}(U)$ suffices, except for Section \ref{wave-gauge_sec} where $\bar{g} \in C^{2,1}(U)$ suffices.} in $U$, and 
\begin{eqnarray}\label{nonzeroderiv}
\big| {\bar{g}}_{\alpha\beta,\gamma} \big| > M \ \ \ \ \text{in} \  U,
\end{eqnarray} 
and, as a consequence of the Lipschitz continuity of $\bar{g}$, 
\begin{eqnarray}\label{linear-dropping}
\big| \bar{g}_{\alpha\beta}(q) \big| \leq \bar{M}\, \delta(q) \ \ \ \ \text{in} \  U,
\end{eqnarray}
where $\delta$ is defined in (\ref{deltaq}).
\end{Def}

Condition (\ref{nonzeroderiv}) says that the shock wave interaction at $p$ is sufficiently localized that there still exists an open set $U$ with $p$ in its closure in which the metric is smooth, but the derivative of the metric has a lower bound within this set, because the metric derivative takes a minimum jump on some shock wave in the neighborhood. This is a very weak structural condition which should easily be met at a point of finite shock wave interaction, including the shock waves which have been simulated in \cite{VoglerTemple}, by choosing $U$ as one of the regions between two adjacent shock curves intersecting in $p$. The metric smoothness in $U$ makes our analysis in Section \ref{sec: perturbed EFE} feasible. 

Our next theorem, gives a precise sense in which approximate locally inertial frames are the closest you can get to an actual locally inertial frame within the smooth atlas in a neighborhood of a point on a smooth shock surface across which the metric is only Lipschitz continuous.   The result extends from single shock surfaces to points of multiple shock wave interaction.

\begin{Thm} \label{Thm_app}  
Assume a gravitational metric $g_{\mu\nu}$ is Lipschitz continuous across a smooth co-dimension one (shock) surface $\Sigma$ in a given coordinate system $x^\mu$, in the sense that the metric is smooth (at least $C^{1,1}$) away from $\Sigma$ and smooth tangential to $\Sigma$, and such that there exists a constant $M>0$ and indices $\mu,\nu,\sigma$ such that the jump in the derivative, $[g_{\mu\nu,\sigma}]$, across the shock surface satisfies
\begin{eqnarray} \label{Thm_app_ineq1}
\big|[g_{\mu\nu,\sigma}]\big|>M .
\end{eqnarray}
Then the following holds: 
(i) For any coordinate system $x^\alpha$ that can be reached within the $C^{2,1}$ atlas, the transformed metric $g_{\alpha\beta}=J^\mu_\alpha J^\nu_\beta g_{\mu\nu}$ is Lipschitz continuous but no smoother and there exist indices $\alpha, \beta,\gamma$ such that across the shock surface 
\begin{eqnarray} \label{Thm_app_ineq2} 
\big|[g_{\alpha\beta,\gamma}](q)\big| >  \|J^{-1}(q) \|^{-3} \, M , 
\end{eqnarray} 
for all $q\in \Sigma$ inside the coordinate patch and where $\|\cdot\|$ denotes the operator norm induced by the maximum norm on $\R^4$, (c.f.  \eqref{norm} below).
(ii) Given a point $p\in\Sigma$, you can always find a coordinate transformation within the $C^{2,1}$ atlas such that $g_{\alpha\beta}(p)=\eta_{\alpha\beta}$, and every such coordinate system is an approximate locally inertial coordinate system in the sense of Definition \ref{defapp}, but never exactly locally inertial.
\end{Thm}

The condition (\ref{Thm_app_ineq2}) shows that $M$ is a uniform bound over all coordinate transformations with bounded Jacobian.  Thus assuming the shock strength is on the order of the jump in the derivatives in the original coordinates, $M$ is then on the order of the shock strength, giving it an invariant physical meaning.   

\Proof
To prove (i), we follow the idea leading to the \emph{smoothing condition}, first introduced in \cite{ReintjesTemple}, which lies at the heart of the method in \cite{Reintjes,ReintjesTemple2}: The covariant transformation law of the metric is given by $g_{\alpha\beta}= J^\mu_\alpha J^\nu_\beta g_{\mu\nu}$, where $J^\mu_\alpha=\frac{\partial x^\mu}{\partial x^\alpha}$ is the Jacobian of the coordinate transformation. Differentiating this transformation law with respect to $\sigma$ (we assume the index $\sigma$ belongs to the coordinates $x^\mu$) and taking the jump of the resulting expression across $\Sigma$ leads to 
\beq \nonumber
[g_{\alpha\beta,\sigma}]= 
\big( [J^\mu_{\alpha,\sigma}] J^\nu_\beta  + J^\mu_\alpha [J^\nu_{\beta,\sigma}] \big) g_{\mu\nu} + J^\mu_\alpha J^\nu_\beta [g_{\mu\nu,\sigma}] ,
\eeq
for all indices $\alpha,\beta$ and $\sigma $. The above equation holds point-wise on $\Sigma$ and the first term on the right hand side vanishes since the Jacobians are continuously differentiable. Thus, contracting the resulting equation with the inverse Jacobians $J^\alpha_\mu$ and $J^\beta_\nu$ gives
\beq \nonumber
J^\alpha_\mu J^\beta_\nu J^\gamma_\sigma [g_{\alpha\beta,\gamma}] =  [g_{\mu\nu,\sigma}] .
\eeq
Taking the maximum of the absolute value of all components of the previous equation, we get
\beq \label{Thm_app_pf_eqn1}
\max_{\mu,\nu,\sigma} \big|  J^\alpha_\mu J^\beta_\nu J^\gamma_\sigma [g_{\alpha\beta,\gamma}] \big| = \max_{\mu,\nu,\sigma} \big| [g_{\mu\nu,\sigma}] \big|> M ,
\eeq
where the lower bound of the right hand side follows from \eqref{Thm_app_ineq1}. The left hand side can now be bounded from above as follows: Let $\|\cdot\|$ denote the induced operator norm (taken point-wise for $q \in \Sigma$), that is, for the inverse Jacobian, $(J^{-1})^\alpha_\mu \equiv J^\alpha_\mu$, we have 
\beq \label{norm}
\|J^{-1}(q)\|= \sup \Big\{ \max\limits_{\mu}|J^\alpha_\mu(q) w_\alpha|\ \Big| \ w_\alpha \in \R, \alpha\in \{0,...,3\} \, \text{and} \, \max_{\alpha}|w_\alpha|=1 \Big\},
\eeq
from which, for fixed $\mu,\nu,\sigma$, we successively get 
\begin{eqnarray} \nonumber
 \big|  J^\alpha_\mu J^\beta_\nu J^\gamma_\sigma [g_{\alpha\beta,\gamma}] \big| 
 &\leq & \big\|  J^{-1} \big\| \cdot \max\limits_{\alpha} \big| J^\beta_\nu J^\gamma_\sigma [g_{\alpha\beta,\gamma}] \big|   \cr
 &\leq & \big\|  J^{-1} \big\|^2 \cdot \max\limits_{\alpha,\beta} \big| J^\gamma_\sigma [g_{\alpha\beta,\gamma}] \big|   \cr
 &\leq & \big\|  J^{-1} \big\|^3 \cdot \max\limits_{\alpha,\beta,\gamma} \big| [g_{\alpha\beta,\gamma}] \big|.
\end{eqnarray}
Using the above inequality to replace the left hand side in \eqref{Thm_app_pf_eqn1}, we obtain
\beq \nonumber
\max\limits_{\alpha,\beta,\gamma} \big| [g_{\alpha\beta,\gamma}] \big| > \big\|  J^{-1} \big\|^{-3} M ,
\eeq
which proves \eqref{Thm_app_ineq2} for some indices $\alpha,\beta,\gamma$, completing the proof of (i). 

We now prove (ii). By the symmetry of $g_{\mu\nu}(p)$ there exist an orthonormal basis of eigenvectors at $p$. Choosing the Jacobian on a neighborhood around $p$ to be the constant matrix that maps the coordinate basis of $x^\mu$ to this orthonormal basis at $p$, we achieve a coordinate system $x^\alpha$ within the smooth atlas for which $g_{\alpha\beta}(p)=\eta_{\alpha\beta}$. We now prove that $x^\alpha$ is an approximate locally inertial frame. Defining $\bar{g}=g-\eta$, the Lipschitz continuity implies \eqref{linear-dropping} with $\bar{M}$ being the Lipschitz constant of $g$ at $p$. The inequality \eqref{nonzeroderiv} follows from the metric smoothness away from the shock surface together with  \eqref{Thm_app_ineq2}, where we can take $U$ to be the open set to the left or right of $\Sigma$. This completes the proof of the Theorem \ref{Thm_app}.
\QED

\section{Regularity Singularities}   \label{sec: regularity singularities}

Assume a Lipschitz continuous gravitational metric whose Riemann curvature tensor is bounded with discontinuities containing no delta function sources, c.f. \cite{ReintjesTemple}.  
\begin{Def}   
We say that a point $p$ in space-time is a {\it regularity singularity} if there does not exist a locally inertial coordinate system at $p$ within the $C^{1,1}$ atlas.   We say a point $p$ is a {\it weak regularity singularity} if there exists locally inertial frames at $p$, but the metric is not $C^{1,1}$ regular in a neighborhood of $p$,  in any coordinate system of the $C^{1,1}$ atlas.
\end{Def}
We conjecture that if a regularity singularity exists, then the $C^{1,1}$ atlas would offer no improvement to the smooth atlas regarding locally inertial frames, \cite{GroahTemple,Reintjes,SmollerTemple}.   Theorem \ref{Thm_app} implies that regularity singularities exist when the atlas is restricted to the class of smooth coordinate transformations. The open question, then,  is whether weak or strong regularity singularities exist with respect to the $C^{1,1}$ atlas, at points of shock wave interaction.  Note that if a metric can be smoothed to $C^{1,1}$ in a neighborhood of $p$, it must then admit locally inertial frames at $p$, but it is an open problem as to whether a metric might admit locally inertial frames at $p$, but not be smoothable to $C^{1,1}$ in a neighborhood of $p$.  Thus we make the distinction between regularity singularities and weak regularity singularities.

\section{Coriolis Type Effects in Approximate Locally Inertial Coordinates and Statement of the Main Result} \label{sec: Coriolis type effects}

In a locally inertial coordinate system centered at a point $p$ where the spacetime is locally flat,  the motion of test particles in free-fall will follow geodesics which are straight lines to errors quadratic in coordinate distance to $p$, because the Christoffel symbols vanish at $p$, i.e., $\Gamma^{l}_{ij}(p)=0$.   Thus, neglecting these second-order errors, the geodesic equation is,
$$
\frac{d^2x^l}{d\tau^2}=\Gamma^{l}_{ij}(p)\dot{x}^i\dot{x}^j=0,
$$
where $\tau$ parameterizes the curve, c.f. \cite{Weinberg}. In contrast, if $p$ is a regularity singularity, then by \eqref{nonzeroderiv}, in each approximate locally inertial coordinate frame, there exists components of the metric whose derivatives are bounded from below by a constant $M>0$ in an open set $U$ containing $p$ in its closure.  Since we can always solve for the derivatives of the metric in terms of the Christoffel symbols at a point via the formula,
$$
\bar{g}_{\alpha\beta,\gamma} = {g}_{\alpha\beta,\gamma} ={\Gamma}_{\alpha\, \beta\gamma}+{\Gamma}_{\beta\,\gamma\alpha},
$$
there exists some set of indices, which for simplicity we again label as $\alpha,\beta,\gamma$, such that
\begin{eqnarray}\label{OneGammabig}
|{\Gamma}^{\gamma}_{\alpha\beta}|\geq M/2,
\end{eqnarray}
in the open set $U$.  
To avoid technicalities, (possible cancellations), assume for simplicity that $\alpha = 0 = \beta$. Then, choosing initial data for the geodesic such that $\dot{x}^0=1$ and $\dot{x}^i=0$ for $i=1,...,3$, the corresponding geodesic equation for the fixed indices $\alpha=0=\beta$ and $\gamma$ becomes initially
$$
\frac{d^2x^{\gamma}}{d\tau^2}   =  \Gamma^{\gamma}_{00},
$$
thereby isolating the presence of an acceleration in the geodesic on the order of $\big|\Gamma^{\gamma}_{00} \big| \geq M/2$ throughout an open set $U$ in each approximate locally inertial coordinate frame. We refer to the non-vanishing terms $\Gamma^{\gamma}_{\alpha\beta}$ as \emph{Coriolis} terms. Since by Israel's theorem points on single shock surfaces admit locally inertial coordinates, such Coriolis terms and the resulting acceleration effects can be removed at points $p$ on single shock surfaces, but by definition such effects could not be removed if $p$ were a regularity singularity.  

Thus, at a regularity singularity, the non-removable Coriolis terms create geodesic deflection and scattering effects which (by definition) could not be removed by coordinate transformations. That is, in case spacetime is locally inertial, the trajectory of a particle following a geodesic curve can locally be approximated by a straight line with second-order error terms (in coordinate distance), which are determined by the curvature of spacetime.  In case a regularity singularity is present at $p$, (larger) \emph{first} order errors, which are due to the Coriolis terms, deflect the possible motion of the trajectory beyond what would be expected by the second-order errors (and hence curvature) alone. Since the Coriolis terms are of order $M$,  the order of the shock strengths,  they would represent physical effects at a regularity singularity should one exist.   

The purpose of the remainder of this paper is to characterize the non-removable Coriolis acceleration terms that appear in the linearized Einstein equations associated with a given approximate locally inertial coordinate system in the sense of Definition \ref{defapp}. In particular, shock wave interactions will typically form in dense matter where geodesic motion of test particles would be highly obscured by the matter, but, gravity waves are weakly interacting with matter, and the scattering of such waves at regularity singularities could be an effect that would in principle be measurable.    In the next section we derive a canonical form for the linearized Einstein equations for gravity waves in approximate locally inertial coordinates, accomplished in Theorem \ref{Prop1}. The main technical problem is to identify non-zero terms first-order in the derivatives of $g$, which survive in the linearized equations after all cancellation is accounted for. Indeed, since the curvature tensor has no delta function sources, we know that delta function sources in the second derivatives must all cancel out in the curvature tensor in an approximate locally inertial coordinate frame.  This then begs the question as to whether there is a corresponding cancellation in the gravity wave equations that makes it so that nonzero first-order derivatives have no physical observable effect due to cancellation as well. We answer this question in Theorem \ref{thmclk} by proving that no such cancellation occurs. Combining Theorems \ref{Prop1} and \ref{thmclk}, we obtain our main theorem which states that no such cancellation occurs, and the linearized Einstein equations cannot be reduced to the Minkowski wave equation at $p$ in any approximate locally inertial frame.  (We always assume a background metric with curvature tensor bounded and free of delta functions sources.)

\begin{Thm}\label{Thmintro}
For each approximate locally inertial coordinate system centered at a point $p$, and for each perturbation of the gravitational metric of the form $\eta + \bar{g} + \epsilon h$, there exist a wave gauge such that, to leading order in $\epsilon>0$, the linearized Einstein equations take the form 
\beq\label{wave eqn intro first}
\frac12 \Box_\eta h_{jl} 
=  \kappa \left(\tilde{T}_{jl} - \frac12 \eta_{jl} \eta^{\sigma\rho} \tilde{T}_{\sigma\rho} \right) 
- \frac{\kappa}{2} \left( h_{jl} \eta^{\sigma\rho} + \eta_{jl} h^{\sigma\rho}  \right) \hat{T}_{\sigma\rho}- h^{ik}\hat{R}_{ijkl}- {{\mathcal C}}_{jl}(h),
\eeq
where $ \Box_\eta\equiv\eta^{\rho\sigma}\frac{\partial^2}{\partial x^{\rho} \partial x^{\sigma}}$ is the wave operator, $\hat{T}$ is the matter source and $\hat{R}$ is the bounded curvature for the background metric $\hat{g}=\eta+\bar{g}$, $\tilde{T}$ is the contribution to the matter sources due to the perturbation $h$, and the terms $C_{ij}$ are the new {\it Coriolis} terms which involve first derivatives of $\bar{g}$ and hence would vanish in an actual locally inertial frame, were one to exist. Moreover, for each locally inertial frame, equations (\ref{wave eqn intro first}) always admit solutions $h$ and indicies $j$ and $l$ such that
\begin{eqnarray}
|\mathcal{C}_{jl}(h)|\geq \frac{1}{4}M
\end{eqnarray}
in some open set containing $p$ in its closure.\footnote{That is, the regularity singularity creates non-removable accelerations in each approximate locally inertial coordinate system within the class of gravity waves that meet the wave gauge in that coordinate system, a physically verifiable condition.}
\end{Thm}

Since a gravity wave at a locally inertial point of spacetime evolves to leading order by the pure wave equation of Minkowski spacetime, the term $\mathcal{C}(h)$ supplies the accelerations which characterize the scattering of gravity waves by the singularity.   Since $\mathcal{C}(h)$ depends on the wave $h$ as well as on the choice of coordinate system, the challenge is to prove that in {\it every} approximate locally inertial frame, there exists an $h$ where the magnitude of $\mathcal{C}$ is on the order of $h$, and this is the substance of the Theorem.   It establishes that regularity singularities would create Coriolis type accelerations that can not be removed by coordinate transformation due to the non-existence of locally inertial frames.   We thus propose that if regularity singularities exist, this is a physical scattering effect of the gravitational field due essentially to the lack of regularity in the underlying spacetime geometry.

\section{Linearizing the Einstein Equations at a regularity singularity}\label{sec: perturbed EFE}\label{linearize}

In this section we formally derive the linearized Einstein equations in approximate locally inertial coordinate systems as defined in Definition \ref{defapp}.   The main issue here is to incorporate the first-order derivatives of the metric into the linearized equations, (c.f. \cite{Schutz} for a detailed description of the linearization procedure around the Minkowski metric).  

To start, assume an approximate locally inertial coordinate system $x^j$ at $p$ where 
\beq\label{background metric Intro}
\hat{g}_{ij}= \eta_{ij} + \bar{g}_{ij},
\eeq
where $\bar{g}$ is at best Lipschitz continuous, and we always assume the curvature tensor is bounded and free of delta function sources.  Gravitational waves propagating through this background spacetime are represented by a symmetric tensor $h_{ij}$ such that the perturbed spacetime metric takes the form
\beq\label{physical metric Intro}
g_{ij}= \eta_{ij} + \bar{g}_{ij} + \epsilon h_{ij},
\eeq
for some sufficiently small constant $\epsilon > 0$. 
In the next paragraph we introduce the wave gauge, and linearize the equations in this gauge.   This is accomplished formally by substituting (\ref{physical metric Intro}) into the Einstein equations in the wave gauge, assuming the neighborhood size $\delta$ is order $\epsilon$,  and discarding terms of order $\epsilon^2$.  The linearized equations (\ref{wave eqn intro first}) then emerge as the equations, linear in $h$, which result at order $\epsilon$.  We then prove that not all $C_{jl}$ can vanish in a given locally inertial coordinate system, and since the equations are linear in $h$, we conclude that there exist solutions $h$ which experience nonzero accelerations due to the Coriolis terms $C_{jl}$.  In this argument we do not address the issue of the validity of neglecting the $O(\epsilon^2)$ terms based on the smoothness of the perturbations $h$.  For example, in the case of two interacting shocks, a gravity wave perturbation $\epsilon h$ would create a small change in the positions of the shock waves encoded in the metric $\bar{g}$, and in such a case the derivatives of the perturbation $\epsilon h$ would not necessarily be small.  However, as long as we can restrict to a subset of the original neighborhood in which the perturbation $h$ is smooth, the linearization procedure would be valid in this neighborhood, and the effect of the Coriolis terms on $h$ would be observed in that neighborhood.   Thus, for our purposes here, we are content with applying the formal linearization procedure and not concerning ourselves with the actual regularity of the perturbations $h$.         
       In line with this, we here understand second order derivatives of the background metric tensor $\bar{g}_{ij}$, which is $C^{1,1}$ regular in the region under consideration, in a point-wise almost everywhere sense.

\subsection{The Wave Gauge} \label{wave-gauge_sec}

To linearize the Einstein equations in approximate locally inertial coordinates,  we introduce the wave gauge condition, (also called the {\it harmonic} gauge condition, \cite{Choquet,HawkingEllis, Schutz}), on solutions of the Einstein equations. The wave gauge condition removes the gauge freedom from the Einstein equations by reducing the leading order terms to the wave operator (i.e., the D'Alambertian), and provides a canonical form for the Coriolis terms in the context of gravitational radiation. 
Differently from the standard linearization procedure, we start by introducing a wave gauge condition on the initial data for $h$ which is propagated {\it exactly} by the full nonlinear Einstein equations, rather than by the linearized equation. In this gauge, the Einstein equations in the highest order derivatives produce the hyperbolic wave operator 
$\left(\eta^{ij}+\bar{g}^{ij}\right)\frac{\partial}{\partial x^i}\frac{\partial}{\partial x^j}$ of the background spacetime.  
Now, $\bar{g}$ is $O(\delta)$ and $\delta=O(\epsilon)$ so only $\eta^{ij}\frac{\partial}{\partial x^i}\frac{\partial}{\partial x^j}$ survives in the leading order derivatives at order $O(\epsilon)$ in the linearization procedure.   The main point then is that derivatives of $\bar{g}$ are $O(1)$ in approximate locally inertial coordinates,  so they appear at order $O(\epsilon)$ on lower order derivatives of the equations in wave gauge, the result being that the first-order derivatives of $\bar{g}$ only affect the lower order derivatives of the resulting linearized equations at order $O(\epsilon)$.  These terms then constitute the sought after Coriolis terms. In particular, by this choice of wave gauge, it follows that we can only expect the wave gauge condition to be maintained within $O(\epsilon^2)$ errors  under evolution by the linearized equations.

Following the development in \cite{Choquet}, we define the wave gauge condition by
\beq\label{wave gauge non-linear}
\omega_i  =0,
\eeq
where 
\beq\label{wave gauge non-linear-1}
\omega_i \equiv \epsilon\, g^{\sigma\rho} \tilde{\Gamma}_{i\, \sigma\rho} ,
\eeq
for
$$
\tilde{\Gamma}_{k\,ij} =  \frac{1}{2} \left( h_{ik,j} + h_{jk,i} - h_{ij,k} \right),
$$ 
so that
\beq\nonumber
\epsilon \tilde{\Gamma}_{k\,ij} =  \Gamma_{k\, ij} - \hat{\Gamma}_{k\, ij} ,
\eeq
with $\Gamma_{k\, ij}$ denoting the Christoffel symbols of $g=\hat{g} +\epsilon h$ and $\hat{\Gamma}_{k\, ij}$ the Christoffel symbols of the background metric $\hat{g}= \eta + \bar{g}$.  
As was shown in Choquet-Bruhat's pioneering work on the existence theory of the Einstein equations, \eqref{wave gauge non-linear} is a condition on the Cauchy data only \cite{Choquet,HawkingEllis}. That is, Choquet-Bruhat proved that whenever the initial data of a solution of the Einstein equations satisfies \eqref{wave gauge non-linear}, the Bianchi identities ensure that \eqref{wave gauge non-linear} holds in the whole Cauchy development. We explain this remarkable property of the Einstein equations in the following formal exposition.  

In her celebrated existence theory, to overcome the problem of the Einstein equations being degenerate hyperbolic, due to the constraint equations, Choquet-Bruhat's strategy was to introduce the \emph{reduced Einstein equations}, a modified hyperbolic version of the Einstein equations to which the Leray-existence theory can be applied directly. The idea is that the solution of the reduced Einstein equations then solves the Einstein equations as long as it satisfies the wave gauge. For our purposes we take the reduced Einstein equations to be
\begin{eqnarray} 
H_{ij}[g] &=& \kappa \left( T_{ij} - \frac12 g_{ij} T_{\sigma\rho}g^{\sigma\rho} \right),  \label{EFE_reduced} \\
{Div}_g T &=& 0, \label{conservation_energy}
\end{eqnarray}
where in a given coordinate system $x^j$ the ``reduced'' Ricci tensor is defined in terms of the Ricci tensor and the gauge fields by
\beq\label{def_reduced_Einstein_tensor}
H_{ij}[g] = R_{ij}[g] + \omega_{i,j} + \omega_{j,i}\, ,
\eeq 
and \eqref{conservation_energy} expresses conservation of energy for the matter fields, which no longer is a consequence of the coupling to the Einstein tensor, c.f. \cite{Choquet,HawkingEllis}. 

The system \eqref{EFE_reduced}-\eqref{conservation_energy} is a system for the pair $(g_{ij},T_{ij})$, giving 14 equations in 14 unknowns.  To compare, the original Einstein equations consist of four constraint equations and six evolutionary equations for $g_{ij}$. Assuming the constraint equations are satisfied initially, and assuming the six evolutionary equations hold, one can use the Bianchi identities to replace the constraint equation by ${Div}_g T=0$, resulting in ten evolutionary equations in 14 unknowns. This leaves four degrees of freedom upon which to impose gauge conditions. The idea of the reduced Einstein equations is to give up the gauge freedom. Namely, fix any coordinate system $x^i$, then \eqref{EFE_reduced} and \eqref{conservation_energy} give 14 equations in 14 unknowns in  $x^i$. The ten equations for the metric are second-order hyperbolic and the four for conservation are first-order hyperbolic, and there is a unique evolution of initial data. We show in the following that this evolution agrees with the evolution of the Einstein equations whenever the gauge condition \eqref{wave gauge non-linear} holds. 

To start, we first show that $H[g]$ is a hyperbolic operator. For this, it suffices to consider the terms of the Ricci tensor containing second-order derivatives. By \eqref{Riemann_prelim} - \eqref{Chr_sym_prelim}, these terms are given by
\begin{eqnarray}\label{E1}
g^{ik}\Gamma_{ij[k,l]}
&=&\frac{1}{2}g^{ik}\left(g_{ik,jl}-g_{jk,il}-g_{il,jk}+g_{jl,ik}\right)\\\nonumber
&=&\frac{1}{2}g^{ik}g_{jl,ik}-\frac{1}{2}\left(g^{ik}\left[g_{jk,il}-\frac{1}{2}g_{ik,jl}\right]+g^{ik}\left[g_{il,jk}-\frac{1}{2}g_{ik,jl}\right]\right)\\\nonumber
&=&\frac{1}{2}\Box_g g_{jl}-  \frac{1}{2}\left(g^{ik} \partial_l \left[g_{jk,i}-\frac{1}{2}g_{ik,j}\right] + g^{ik} \partial_j \left[g_{il,k}-\frac{1}{2}g_{ik,l}\right]\right), \nonumber
\end{eqnarray}
where $\Box_g g_{jl} \equiv g^{ik}g_{jl,ik}$ denotes the wave equation based on $g$ and 
\beq \nonumber
\Gamma_{ij[k,l]} \equiv \Gamma_{ijk,l} - \Gamma_{ijl,k} . 
\eeq
A straight-forward computation starting from the definition of the Christoffel symbols \eqref{Chr_sym_prelim} shows that 
$$ 
g^{ik}\left( g_{jk,i}- \frac{1}{2}g_{ik,j}\right) \ =\ g^{\sigma\rho} {\Gamma}_{j\, \sigma\rho} \ =\  g^{\sigma\rho} \left( \hat{\Gamma}_{j\, \sigma\rho} + \epsilon \tilde{\Gamma}_{j\, \sigma\rho} \right) ,
$$ 
where we used that $\Gamma_{k\, ij} =  \hat{\Gamma}_{k\, ij} + \epsilon\, \tilde{\Gamma}_{k\, ij}$. We now conclude by \eqref{wave gauge non-linear-1} that
\beq\label{gauge_function_eqn1}
\omega_j = g^{ik}\left( g_{jk,i}- \frac{1}{2}g_{ik,j}\right) - g^{ik} \hat{\Gamma} _{j\, ik}.
\eeq
Substituting \eqref{gauge_function_eqn1} into \eqref{E1} gives the leading order part of the Ricci tensor as
\begin{eqnarray}\label{E2}
g^{ik}\Gamma_{ij[k,l]}
&=&\frac{1}{2}\Box_g g_{jl}-\left(\omega_{j,l}+\omega_{l,j}\right)  + l.o.t. \cr
&=&\frac{\epsilon}{2}\Box_g h_{jl}-\left(\omega_{j,l}+\omega_{l,j}\right)  + l.o.t.,
\end{eqnarray} 
where $l.o.t.$ denotes sums of all terms not relevant to the hyperbolic structure of the Einstein equations for $h$, that is, sums of terms containing $h$ and first derivatives of $h$, together with terms containing up to second-order derivatives of the background metric $\hat{g}$.  From \eqref{E2}, it follows that
\beq\nonumber
R_{ij} = \frac{\epsilon}{2}\Box_g h_{jl}-\left(\omega_{j,l}+\omega_{l,j}\right)  + l.o.t.
\eeq 
and hence 
\beq \nonumber
H_{ij}[g] = \frac{\epsilon}{2}\Box_g h_{jl} + l.o.t. \, .
\eeq
The conclusion then is that the reduced Einstein equations are hyperbolic and reduce to the Einstein equations whenever the gauge condition \eqref{wave gauge non-linear} holds, c.f. \cite{Choquet}.

In this context, we now outline Choquet-Bruhat's argument for the wave gauge \eqref{wave gauge non-linear} being propagated by the reduced Einstein equations, \eqref{EFE_reduced} - \eqref{conservation_energy}. That is, we show that if the wave gauge holds initially, then it is satisfied in the whole Cauchy development. To begin, assume the Einstein constraint equations and the wave gauge conditions are satisfied by the initial data. In order to apply the Bianchi identities in the form $\text{Div}_g G=0$, we write \eqref{EFE_reduced} in its equivalent form
\beq\label{EFE_reduced2}
H_{ij}[g] - \frac12 g_{ij} H_{\sigma\rho}[g]g^{\sigma\rho} = \kappa  T_{ij} ,
\eeq
and correspondingly the definition of $H_{ij}[g]$ in \eqref{def_reduced_Einstein_tensor} as
\begin{eqnarray} \label{def_reduced_Einstein_tensor2}
H_{ij}[g] - \frac12 g_{ij} H_{\sigma\rho}[g]g^{\sigma\rho} 
&=& G_{ij}[g] + \omega_{i,j} + \omega_{j,i} - \frac12 g_{ij} \left(  \omega_{\sigma,\rho} + \omega_{\rho,\sigma} \right) g^{\sigma\rho} \cr
&=& G_{ij}[g] + \omega_{i,j} + \omega_{j,i} -  g_{ij} g^{\sigma\rho} \omega_{\sigma,\rho}  . 
\end{eqnarray}
Then substituting \eqref{def_reduced_Einstein_tensor2} into \eqref{EFE_reduced2} leads to the following equivalent form of the reduced Einstein equations
\beq\label{E3-0}
\kappa T_{ij} - G_{ij}[g] = \omega_{i,j} + \omega_{j,i}  -  g_{ij} g^{\sigma\rho} \omega_{\sigma,\rho}.
\eeq 
Taking the divergence of \eqref{E3-0} and using the conservation equation \eqref{conservation_energy} as well as the Bianchi identities in the form $\text{Div}_g G=0$, the left hand side of \eqref{E3-0} vanishes and we obtain
\beq\label{E3}
g^{j\tau}\omega_{i,j\tau} + g^{j\tau}\omega_{j,i\tau}
-  g^{\sigma\rho} \omega_{\sigma,\rho i}  
+ L_i^{\sigma\rho} \omega_{\sigma,\rho}   = 0,
\eeq 
where $L_i^\sigma$ depends only on $g$ and on the Christoffel symbols of $g$. A direct computation shows that the second and third terms in \eqref{E3} cancel, so that \eqref{E3} is equivalent to
\beq\label{E4}
g^{\sigma\rho} \omega_{i,\sigma\rho} + L_i^{\sigma\rho} \omega_{\sigma,\rho}   = 0.
\eeq
The main point now is that \eqref{E4} is a {\it homogeneous} hyperbolic second-order equation for $\omega_i$, so that the unique solution of \eqref{E4} vanishes whenever its initial data vanishes. Thus, to prove that the wave gauge holds in the Cauchy development, assuming the gauge condition \eqref{wave gauge non-linear} holds initially, it remains only to prove that the derivatives of $\omega_i$ in a direction normal to the Cauchy surface vanish initially. For our purposes it suffices to consider the Cauchy surface $\Sigma = \{ x^0=0\}$ only. We now show that the sought-after vanishing of $\omega_{i,0}$ on $\Sigma$ follows from the Einstein constraint equations on $\Sigma$. 

The Einstein constraint equations are given on the surface $\Sigma $ by 
\beq\label{constraint-eqn}
G_{i0}[g] = \kappa T_{i0}.
\eeq
These are precisely the four Einstein equations for which the second-order $x^0$-derivatives drop out. Substituting \eqref{EFE_reduced2} for the right hand side of \eqref{constraint-eqn} gives us
\beq\label{E5}
G_{i0}[g] = H_{i0}[g] - \frac12 g_{i0} H_{\sigma\rho}[g]g^{\sigma\rho}
\eeq
on $\Sigma$. By \eqref{def_reduced_Einstein_tensor2}, equations \eqref{E5} are equivalent to
\beq\label{E6}
\omega_{i,0} + \omega_{0,i} = g_{i0} g^{\sigma\rho} \omega_{\sigma,\rho}  ,
\eeq
and contraction with $g^{ij}$ results in
\beq\label{E6_b}
g^{ij} \left(\omega_{i,0} + \omega_{0,i} \right)= \delta^j_{0} g^{\sigma\rho} \omega_{\sigma,\rho}  ,
\eeq
where $\delta^j_i$ denotes the Kronecker delta. Now, the wave gauge \eqref{wave gauge non-linear} implies that $\omega_{i,j}=0$ for $j\neq0$ everywhere on $\Sigma$, and from this we find that \eqref{E6_b} is equivalent to
\beq\label{E6_c}
g^{\sigma j} \omega_{\sigma,0} + g^{0j} \omega_{0,0} - \delta^j_0\: g^{\sigma 0} \omega_{\sigma,0} = 0.
\eeq
In the context of approximate locally inertial frames $g^{00}$ can always be taken to be non-zero, thus the $j=0$ component of \eqref{E6_c} implies $w_{0,0}=0$ on $\Sigma$. This then implies that the $j=\alpha$ component, for $\alpha \in \{1,2,3\}$, is given by
\beq \nonumber
g^{\alpha\sigma} \omega_{\sigma,0} =0,
\eeq
and since the metric is non-singular we finally obtain
\beq\label{E7}
\omega_{i,0}  = 0
\eeq
on $\Sigma$ for all $i\in \{0,...,3\}$. This proves that \eqref{E7} holds whenever \eqref{wave gauge non-linear} holds initially and therefore the wave gauge is propagated by the reduced Einstein equations.

\subsection{The Linearized Einstein Equations in Approximate Locally Inertial Frames}

In the above subsection, we have shown that the wave gauge is propagated by solutions of the Einstein equations \eqref{EFE_reduced}-\eqref{conservation_energy}, as long as the wave gauge and the Einstein constraint equations hold initially. Based on this, we assume the wave gauge \eqref{wave gauge non-linear} at the start, and derive the linearized Einstein equations in wave gauge in an approximate locally inertial frame.   This formal procedure accomplishes the following theorem, which identifies the non-removable Coriolis terms we seek. 

\begin{Thm}\label{Prop1}  
Assume the background metric $\hat{g}_{ij}=\eta_{ij}+\bar{g}_{ij}$ is given in approximate locally inertial coordinates and solves the Einstein equations for a source $\hat{T}$, while its perturbation $g_{ij}=\eta_{ij} + \bar{g}_{ij} + \epsilon h_{ij}$ solves the Einstein equations for some perturbed source $T \equiv \hat{T} + \epsilon \tilde{T}$. In addition, assume that $\bar{g}=O(\delta)$, $\delta=O(\epsilon)$ and $h$ and derivatives of $h$ are $O(1)$ as $\epsilon$ tends to $0$, and assume that the perturbation $h_{ij}$ is in the wave gauge \eqref{wave gauge non-linear}. Then, formally, substituting the ansatz $g_{ij}=\eta_{ij} + \bar{g}_{ij} + \epsilon h_{ij}$ into the Einstein equations, dropping $O(\epsilon^2)$-terms and dividing by $\epsilon$, leads to the following linearized equations for $h$:
\beq\label{EFE phys spacetime 4}
\frac12 \Box_\eta h_{jl} + {{\mathcal C}}_{jl}(h) - h^{ik}\hat{R}_{ijkl}
=  \kappa \left(\tilde{T}_{jl} - \frac12 \eta_{jl} \eta^{\sigma\rho} \tilde{T}_{\sigma\rho} \right) 
- \frac{\kappa}{2} \left( h_{jl} \eta^{\sigma\rho} - \eta_{jl} h^{\sigma\rho}  \right) \hat{T}_{\sigma\rho}.
\eeq
Here $\kappa=-\frac{8\pi}{c^4} \mathcal{G}$,\footnote{Recall that $\kappa$ is negative by our sign convention of the Riemann curvature tensor, as we discussed in a footnote in the Preliminaries. Using the opposite sign convention, the term $- h^{ik}\hat{R}_{ijkl}$ must be replaced by $+ h^{ik}\hat{R}_{ijkl}$ in \eqref{EFE phys spacetime 4} with no other changes required.} $\Box_\eta$ denotes the flat linear wave operator, 
\beq\nonumber
\Box_\eta h_{jl} \equiv \eta^{\sigma\rho}h_{jl,\sigma\rho}
\eeq
and $\mathcal{C}_{jl}(h)$ are given by
\beq\label{def Coriolis term}
\mathcal{C}_{jl}(h) \equiv  {b}_{jl \sigma}^{\rho\tau} \tilde{\Gamma}^\sigma_{\ \rho\tau} - h^{ik} {S}_{ijkl}
\eeq
where
\begin{eqnarray}
{S}_{ijkl} 
&\equiv &  \bar{\Gamma}^\sigma_{\,i l} \bar{\Gamma}_{k\,j\sigma} - \bar{\Gamma}^\sigma_{\,i \sigma} \bar{\Gamma}_{k\,jl},  \label{def S for EFE physical} \\ 
b_{jl\sigma}^{\rho\tau} 
&\equiv &\delta^{\tau}_{l}\bar{\Gamma}^{\rho}_{j\sigma}
-\delta^{\tau}_{\sigma}\bar{\Gamma}^{\rho}_{jl}
+\delta^{\rho}_{j}\bar{\Gamma}^{\tau}_{\sigma l}
-\delta^{\rho}_{j}\delta^{\tau}_{l}\bar{\Gamma}^{k}_{\sigma k}  +  
\frac12\left( \eta_{j\sigma} \hat{g}^{\rho\tau}_{\ \ ,l}   + \eta_{l\sigma} \hat{g}^{\rho\tau}_{\ \ ,j} \right)\hspace{-.1cm}, \ \ \ \ \ \ \ \label{def_b}
\end{eqnarray}
\begin{eqnarray}
\tilde{\Gamma}_{k\,ij} &\equiv & \frac12 \left( h_{ik,j} + h_{jk,i} - h_{ij,k} \right),  \label{Chr sym perturbation} \\
\bar{\Gamma}_{k\,ij} &\equiv & \frac12 \left( \bar{g}_{ik,j} + \bar{g}_{jk,i} - \bar{g}_{ij,k} \right), \label{Chr sym background}
\end{eqnarray}
and we use the notation $h^{ik} \equiv\eta^{i\sigma}\eta^{k\rho} h_{\sigma\rho}$, $\bar{\Gamma}^l_{\,ij}\equiv \eta^{lk}\bar{\Gamma}_{k\,ij}$ and $\tilde{\Gamma}^l_{\,ij}\equiv \eta^{lk} \tilde{\Gamma}_{k\,ij}$ and $\hat{g}^{ij}$ denotes the inverse of $\hat{g}_{ij}$.
\end{Thm}

The terms $\mathcal{C}_{jl}(h)$, the Coriolis terms, are precisely the terms in \eqref{EFE phys spacetime 4} which do not appear in the linearized Einstein equations in an exact locally inertial frame. It is important to note that $\mathcal{C}_{jl}(h)$ contains derivatives of $\bar{g}_{ij}$ which are large throughout the approximate locally inertial frame in a neighborhood of $p$, not just at $p$ itself.            

\Proof  By assumption, the background metric, $\hat{g}_{ij}=\eta_{ij} + \bar{g}_{ij}$, satisfies the Einstein equations for the energy momentum tensor $\hat{T}_{ij}$, that is,
\beq\label{EFE background}
\hat{R}_{jl} \equiv R_{jl}[\hat{g}] = \kappa \left( \hat{T}_{jl} - \frac12 \hat{g}_{jl} \hat{g}^{\sigma\rho} \hat{T}_{\sigma\rho} \right),
\eeq
while the perturbed metric, $g_{ij}=\eta_{ij}+ \bar{g}_{ij} + \epsilon h_{ij}$, solves the Einstein equations for a likewise perturbed energy and matter source, $T_{ij} = \hat{T}_{ij}+ \epsilon \tilde{T}_{ij}$, namely,
\beq\label{EFE phys spacetime}
R_{jl}[\hat{g}+\epsilon h] = \kappa \left( T_{jl} - \frac12 {g}_{jl} {g}^{\sigma\rho} T_{\sigma\rho} \right).
\eeq

For the purpose of raising and lowering indices, we now introduce approximate expressions for the inverse of $g_{ij}$ and of $\hat{g}_{ij}$. To begin with, denote the exact inverse of $g_{ij}$ with $g^{ij}$ and the exact inverse of $\hat{g}_{ij}$ with $\hat{g}^{ij}$. Then, setting 
\beq \nonumber
\bar{g}^{ij}=\eta^{i\sigma} \eta^{j\rho} \bar{g}_{\sigma\rho},
\eeq
we find by cancellation that
\begin{eqnarray} \nonumber
\left( \eta^{i\sigma} - \bar{g}^{i\sigma}\right) \hat{g}_{\sigma j} 
&=& \left( \eta^{i\sigma} - \bar{g}^{i\sigma}\right) \left( \eta_{\sigma j} + \bar{g}_{\sigma j} \right) \cr 
&=& \delta^i_j - \bar{g}^{i\sigma} \bar{g}_{\sigma j} \cr 
&=& \delta^i_j + O(\epsilon^2),
\end{eqnarray}
and since $\hat{g}_{ij}$ is assumed $O(1)$, we conclude that
\beq\label{g_bar upper indices}
\hat{g}^{ij} = \eta^{ij} - \bar{g}^{ij} + O(\epsilon^2).
\eeq
Likewise, defining
\beq \nonumber
h^{ij} \equiv \eta^{i\sigma} \eta^{j\rho} h_{\sigma\rho},
\eeq
a computation using \eqref{g_bar upper indices} gives
$$
\left( \hat{g}^{i\sigma} - \epsilon h^{i\sigma} \right) {g}_{\sigma j}  = \left( \hat{g}^{i\sigma} - \epsilon h^{i\sigma} \right) \left( \hat{g}_{\sigma j} + \epsilon h_{\sigma j} \right) =  \delta^i_j + O(\epsilon^2),
$$
and since ${g}_{ij}$ is assumed $O(1)$, it follows that
\beq\label{inverse perturbed metric}
g^{ij} = \hat{g}^{ij} - \epsilon h^{ij} +O(\epsilon^2).
\eeq 
In the argument below we are careful not to take derivatives of \eqref{inverse perturbed metric} or \eqref{g_bar upper indices}, as this could introduce larger errors in the derivative.

Now, using the metric ansatz \eqref{physical metric Intro} and our approximation of the inverse metric in \eqref{inverse perturbed metric}, we write the Einstein equation \eqref{EFE phys spacetime} as
\beq\label{EFE phys spacetime 2}
R_{jl}[\hat{g}+\epsilon h] = \kappa \left( T_{jl} - \frac12 \hat{g}_{jl} \hat{g}^{\sigma\rho} T_{\sigma\rho} \right) - \frac{\kappa\epsilon}{2} \left( h_{jl} \hat{g}^{\sigma\rho} - \hat{g}_{jl} h^{\sigma\rho}  \right) {T}_{\sigma\rho} + O(\epsilon^2) ,
\eeq
which simplifies after substituting $T_{ij}=\hat{T}_{ij} + \epsilon\, \tilde{T}_{ij}$ and applying \eqref{g_bar upper indices} to the form
\begin{eqnarray}\label{EFE phys spacetime 2a}
R_{jl}[\hat{g} + \epsilon h] &=& \kappa \left( \hat{T}_{jl} - \frac12 \hat{g}_{jl} \hat{g}^{\sigma\rho} \hat{T}_{\sigma\rho} \right) + \kappa\epsilon \left(\tilde{T}_{jl} - \frac12 \eta_{jl} \eta^{\sigma\rho} \tilde{T}_{\sigma\rho} \right) \cr
&&- \frac{\kappa\epsilon}{2} \left( h_{jl} \eta^{\sigma\rho} - \eta_{jl} h^{\sigma\rho}  \right) \hat{T}_{\sigma\rho} + O(\epsilon^2) .
\end{eqnarray}

We now expand the left hand side of \eqref{EFE background} and \eqref{EFE phys spacetime 2a}. Recall that the Ricci tensor is the trace of the Riemann tensor, 
\beq \label{Ricci}
R_{jl}= g^{ik}R_{ijkl},
\eeq
and the Riemann tensor is given in terms of the Christoffel symbols by the formula
\beq\label{Riemann}
R_{ijkl}=  \Gamma_{ij[k,l]} + g^{\sigma\rho} \left( \Gamma_{i \sigma l} \Gamma_{\rho jk} - \Gamma_{i \sigma k} \Gamma_{\rho jl} \right),
\eeq 
where we define the commutator 
\beq\nonumber 
\Gamma_{ij[k,l]}= \Gamma_{ijk,l} - \Gamma_{ijl,k},
\eeq
and 
$$ 
\Gamma_{k\,ij} = \frac12 \left( g_{ik,j} + g_{jk,i} - g_{ij,k} \right).
$$ 
Moreover, due to \eqref{Chr sym perturbation} and \eqref{Chr sym background}, the Christoffel symbols separate as
\beq\label{Chr_sym_splitting}
\Gamma_{k\,ij} = \bar{\Gamma}_{k\,ij} + \epsilon \,\tilde{\Gamma}_{k\,ij}.
\eeq
Now, by \eqref{Ricci} and \eqref{Riemann}, the second-order derivative terms of the Ricci tensor are given by $g^{ik} \Gamma_{ij[k,l]}$, and to separate these into its $\bar{\Gamma}_{ijk}$- and $\tilde{\Gamma}_{ijk}$-dependence, use that \eqref{inverse perturbed metric} agrees with the inverse metric up to $O(\epsilon^2)$ errors, giving
\begin{eqnarray}
g^{ik} \Gamma_{ij[k,l]} 
&=& \left( \hat{g}^{ik} - \epsilon h^{ik} \right) \left( \bar{\Gamma}_{ij[k,l]} + \epsilon \tilde{\Gamma}_{ij[k,l]} \right)+O(\epsilon^2) \nonumber \\
&=& \hat{g}^{ik} \bar{\Gamma}_{ij[k,l]} - \epsilon h^{ik} \bar{\Gamma}_{ij[k,l]} + \epsilon \hat{g}^{ik} \tilde{\Gamma}_{ij[k,l]}+O(\epsilon^2). \label{Ricci 2nd order deriv}
\end{eqnarray}
To expand the lower-order derivative terms in \eqref{Riemann} in $\epsilon$, using that all terms containing $\epsilon\tilde{\Gamma}$ squared are of order $O(\epsilon^2)$, we obtain
\beq\label{Ricci nonlin terms}
g^{ik} g^{\sigma\rho} \left( \Gamma_{i\,\sigma l} \Gamma_{\rho\,jk} - \Gamma_{i\,\sigma k} \Gamma_{\rho\,jl} \right) = \{ \cdot\}_\text{I} + \{ \cdot\}_\text{II} + \{ \cdot\}_\text{III} + O(\epsilon^2),
\eeq
where
\begin{eqnarray}
\{ \cdot\}_\text{I} &\equiv & \hat{g}^{ik} \hat{g}^{\sigma\rho}\left( \bar{\Gamma}_{i\,\sigma l} \bar{\Gamma}_{\rho\,jk} - \bar{\Gamma}_{i\,\sigma k} \bar{\Gamma}_{\rho\,jl} \right) ,\label{Ricci nonlin terms - 1} \\
\{ \cdot\}_\text{II} &\equiv & - \epsilon \left( \hat{g}^{ik} h^{\sigma\rho} + h^{ik} \hat{g}^{\sigma\rho} \right) \left( \bar{\Gamma}_{i\,\sigma l} \bar{\Gamma}_{\rho\,jk} - \bar{\Gamma}_{i\,\sigma k} \bar{\Gamma}_{\rho\,jl} \right), \label{Ricci nonlin terms - 2} \\
\{ \cdot\}_\text{III} &\equiv & \epsilon \hat{g}^{ik} \hat{g}^{\sigma\rho} \left( \tilde{\Gamma}_{i\,\sigma l} \bar{\Gamma}_{\rho\,jk} - \tilde{\Gamma}_{i\,\sigma k} \bar{\Gamma}_{\rho\,jl} + \bar{\Gamma}_{i\,\sigma l} \tilde{\Gamma}_{\rho\,jk} - \bar{\Gamma}_{i\,\sigma k} \tilde{\Gamma}_{\rho\,jl}\right).\ \ \ \label{Ricci nonlin terms - 3} 
\end{eqnarray}
Combining \eqref{Ricci 2nd order deriv} and \eqref{Ricci nonlin terms}, we write the Einstein equations \eqref{EFE phys spacetime 2a} as
\begin{eqnarray}\label{EFE phys spacetime 2b}
 R_{jl}[\hat{g}+\epsilon h] 
&\equiv &   \hat{g}^{ik} \bar{\Gamma}_{ij[k,l]} - \epsilon h^{ik} \bar{\Gamma}_{ij[k,l]} + \epsilon \hat{g}^{ik} \tilde{\Gamma}_{ij[k,l]}  \cr 
& & +   \{ \cdot\}_\text{I} + \{ \cdot\}_\text{II} + \{ \cdot\}_\text{III}  + O(\epsilon^2) \cr
&=& \kappa \left( \hat{T}_{jl} - \frac12 \hat{g}_{jl} \hat{g}^{\sigma\rho} \hat{T}_{\sigma\rho} \right) + \kappa\epsilon \left(\tilde{T}_{jl} - \frac12 \eta_{jl} \eta^{\sigma\rho} \tilde{T}_{\sigma\rho} \right) \cr
&&- \frac{\kappa\epsilon}{2} \left( h_{jl} \eta^{\sigma\rho} - \eta_{jl} h^{\sigma\rho}  \right) \hat{T}_{\sigma\rho}  .
\end{eqnarray}

To simplify \eqref{EFE phys spacetime 2b} further, observe that \eqref{Ricci nonlin terms - 1} and the first term in \eqref{Ricci 2nd order deriv} combine to give the Ricci tensor for the background metric,
 so that we can write the Einstein equations for the background metric \eqref{EFE background} as 
\beq\label{EFE background 2}
R_{jl}[\hat{g}] \equiv \hat{g}^{ik} \bar{\Gamma}_{ij[k,l]} + \{ \cdot\}_\text{I} \ = \ \kappa \left( \hat{T}_{jl} - \frac12 \hat{g}_{jl} \hat{g}^{\sigma\rho} \hat{T}_{\sigma\rho} \right).
\eeq
Now subtracting the Einstein equations for the background metric \eqref{EFE background 2} from the above form of Einstein  equations for the perturbed metric \eqref{EFE phys spacetime 2b}, we obtain 
\begin{eqnarray}\label{EFE phys spacetime 3}
 \epsilon\hat{g}^{ik} \tilde{\Gamma}_{ij[k,l]} &-&\epsilon h^{ik} \bar{\Gamma}_{ij[k,l]} + \{ \cdot\}_\text{II} + \{ \cdot\}_\text{III} + O(\epsilon^2) \cr 
&=&   \kappa\epsilon \left(\tilde{T}_{jl} - \frac12 \hat{g}_{jl} \hat{g}^{\sigma\rho} \tilde{T}_{\sigma\rho} \right) - \frac{\kappa\epsilon}{2} \left( h_{jl} \hat{g}^{\sigma\rho} - \hat{g}_{jl} h^{\sigma\rho}  \right) \hat{T}_{\sigma\rho}. \ \ \ \ \ \
\end{eqnarray}

For further simplification of \eqref{EFE phys spacetime 3}, we use the formula \eqref{Riemann} for the Riemann tensor of the background metric, which leads to 
\begin{eqnarray}
- \epsilon h^{ik} \bar{\Gamma}_{ij[k,l]} + \{ \cdot\}_\text{II} 
&=& - \epsilon h^{ik} \Big( \bar{\Gamma}_{ij[k,l]}  + \hat{g}^{\sigma\rho} \left( \bar{\Gamma}_{i\,\sigma l} \bar{\Gamma}_{\rho\,jk} - \bar{\Gamma}_{i\,\sigma k} \bar{\Gamma}_{\rho\,jl} \right)  \Big) \cr
 & & - \, \epsilon \hat{g}^{ik} h^{\sigma\rho} \left( \bar{\Gamma}_{i\,\sigma l} \bar{\Gamma}_{\rho\,jk} - \bar{\Gamma}_{i\,\sigma k} \bar{\Gamma}_{\rho\,jl} \right)   \cr
&=& - \epsilon h^{ik} \hat{R}_{ijkl} - \epsilon \hat{g}^{ik} h^{\sigma\rho} \left( \bar{\Gamma}_{i\,\sigma l} \bar{\Gamma}_{\rho\,jk} - \bar{\Gamma}_{i\,\sigma k} \bar{\Gamma}_{\rho\,jl} \right)  \cr
&=&  - \epsilon h^{ik} \hat{R}_{ijkl} -  \epsilon h^{ik} {S}_{ijkl}  +   O(\epsilon^2), \label{finaleqn_Prop1_b}
\end{eqnarray}
where we substituted \eqref{g_bar upper indices} for $\hat{g}^{ik}$ in the last equality together with the definition of $S_{ijkl}$, \eqref{def S for EFE physical}. Thus, substituting \eqref{finaleqn_Prop1_b} into \eqref{EFE phys spacetime 3}, the linearized Einstein equations obtain the form
\begin{eqnarray}\label{EFE phys spacetime 3b}
 \epsilon\hat{g}^{ik} \tilde{\Gamma}_{ij[k,l]} &+& \{ \cdot\}_\text{III} - \epsilon h^{ik} \hat{R}_{ijkl} -  \epsilon h^{ik} {S}_{ijkl} + O(\epsilon^2) \cr 
&=&   \kappa\epsilon \left(\tilde{T}_{jl} - \frac12 \hat{g}_{jl} \hat{g}^{\sigma\rho} \tilde{T}_{\sigma\rho} \right) - \frac{\kappa\epsilon}{2} \left( h_{jl} \hat{g}^{\sigma\rho} - \hat{g}_{jl} h^{\sigma\rho}  \right) \hat{T}_{\sigma\rho}. \ \ \ \ \ \
\end{eqnarray}
To derive the final form of the linearized Einstein equations in approximate locally inertial coordinates it remains only to impose the wave gauge condition. The main step is the following lemma.
 
 \begin{Lemma}
 Assume $h_{ij}$ satisfies the wave gauge \eqref{wave gauge non-linear}, then
\beq\label{claim2}
\epsilon\hat{g}^{ik} \tilde{\Gamma}_{ij[k,l]} + \{\cdot\}_\emph{III}   =  \frac\epsilon2 \Box_\eta h_{jl} + \epsilon\, {b}_{jl \sigma}^{\rho\tau}\tilde{\Gamma}^\sigma_{\ \rho\tau} + O(\epsilon^2) ,
\eeq
where ${b}_{jl \sigma}^{\rho\tau}$ is defined in \eqref{def_b}.
\end{Lemma}

Before giving the proof of the Lemma, note that \eqref{claim2} implies the linearized Einstein equations \eqref{EFE phys spacetime 4}, by substituting \eqref{claim2} into \eqref{EFE phys spacetime 3b} and dividing the resulting equation by $\epsilon$, yielding
\begin{eqnarray}\label{EFE phys spacetime 3c}
&&  \frac12 \Box_\eta h_{jl} +  {b}_{jl \sigma}^{\rho\tau}\tilde{\Gamma}^\sigma_{\ \rho\tau}  -  h^{ik} {S}_{ijkl}  - h^{ik} \hat{R}_{ijkl} + O(\epsilon) \cr 
&=&   \kappa \left(\tilde{T}_{jl} - \frac12 \eta_{jl} \eta^{\sigma\rho} \tilde{T}_{\sigma\rho} \right) - \frac{\kappa}{2} \left( h_{jl} \eta^{\sigma\rho} - \eta_{jl} h^{\sigma\rho}  \right) \hat{T}_{\sigma\rho},
\end{eqnarray}
which is \eqref{EFE phys spacetime 4}. To complete the proof Theorem \ref{Prop1}, it remains only to prove the lemma. \vspace{.2cm}

\noindent \underline{Proof of the Lemma:} We first show that using the wave gauge condition, \eqref{wave gauge non-linear}, the second-order derivative terms combine to the flat wave operator $\Box_\eta$. To begin with, substitute \eqref{g_bar upper indices} for $\hat{g}^{ik}$ in the second-order terms in \eqref{EFE phys spacetime 3} and use that $\bar{g}^{ij}=O(\delta)$, we then get
\begin{eqnarray}
\epsilon\hat{g}^{ik} \tilde{\Gamma}_{ij[k,l]} 
&=& \epsilon \left( \eta^{ik} - \bar{g}^{ik} \right)  \tilde{\Gamma}_{ij[k,l]} + O(\epsilon^2) \cr
&=&  \epsilon \eta^{ik} \tilde{\Gamma}_{ij[k,l]} + O(\epsilon^2), \label{Prop1_2derivative-term_eqn1} 
\end{eqnarray}
while a straight forward computation leads to (c.f. \eqref{E1})
\begin{eqnarray}\label{Prop1_2derivative-term_eqn2} 
\eta^{ik} \tilde{\Gamma}_{i j[k,l]} 
&=& \tilde{\Gamma}^k_{\ jk,l} - \tilde{\Gamma}^k_{\ jl,k} \cr
&=& \frac12 \left( h^k_{\ k,jl} - h^k_{\ j,kl} - h^k_{\ l,jk} + \eta^{\sigma\rho} h_{jl,\sigma\rho}  \right) \cr
&=& \frac12 \Box_\eta h_{jl} - \frac12 \partial_l \left( h^k_{\ j,k} - \frac12 h^k_{\ k,j} \right) - \frac12 \partial_j\left( h^k_{\ l,k} - \frac12 h^k_{\ k,l} \right)\hspace{-.15cm}, \ \ \ \ \ \ \ \ 
\end{eqnarray}
where we raise indices on $h_{ij}$ and $\tilde{\Gamma}_{i jk}$ with the Minkowski metric. From the definition of $\tilde{\Gamma}_{i jk}$, \eqref{Chr sym perturbation}, a straight forward computation shows that 
\beq\label{wave_gauge_eqn1 }
 h^k_{\ l,k} - \frac12 h^k_{\ k,l}  = \eta^{\sigma\rho} \tilde{\Gamma}_{l\, \sigma\rho} \ ,
\eeq
and substituting \eqref{wave_gauge_eqn1 } into \eqref{Prop1_2derivative-term_eqn2} yields
\begin{eqnarray}\label{Prop1_2derivative-term_eqn2a} 
\eta^{ik} \tilde{\Gamma}_{i j[k,l]} 
&=& \frac12 \Box_\eta h_{jl} - \frac12 \partial_l \left( \eta^{\sigma\rho} \tilde{\Gamma}_{j\, \sigma\rho} \right) - \frac12 \partial_j  \left( \eta^{\sigma\rho}\tilde{\Gamma}_{l\, \sigma\rho} \right).
\end{eqnarray}

Let us remark at this point, if locally inertial frames existed, we could impose the gauge condition $\eta^{\sigma\rho} \tilde{\Gamma}_{l\, \sigma\rho} =0$, which would be propagated by the resulting flat wave equation, in which case the right hand side of \eqref{Prop1_2derivative-term_eqn2a} would reduce to $\frac12 \Box_\eta$. However, in an approximate locally inertial frame, we work with the wave gauge condition \eqref{wave gauge non-linear} which is propagated by the full non-linear Einstein equation. In order to use \eqref{wave gauge non-linear} in \eqref{Prop1_2derivative-term_eqn2a}, we have to separate $\eta^{\sigma\rho} \tilde{\Gamma}_{l\, \sigma\rho}$ from the remaining terms in the wave gauge condition. Naively, this could be achieved using the approximate inverse of $g_{ij}$ in \eqref{inverse perturbed metric}, then the wave gauge condition assumes the form
\beq\label{wave_gauge_naive}
\eta^{\sigma\rho} \tilde{\Gamma}_{i\, \sigma\rho} = \left( \hat{g}^{\sigma\rho} + \epsilon h^{\sigma\rho} \right) \tilde{\Gamma}_{i\, \sigma\rho} + O(\epsilon^2).
\eeq
However, using \eqref{wave_gauge_naive} in \eqref{Prop1_2derivative-term_eqn2a}, at this stage we would loose control of the error due to the derivatives $\partial_l$ and $\partial_j$ on the gauge functions. Therefore, we approximate $g^{ij}$ in a way such that we can control derivatives of the error terms. For this, we introduce
\begin{eqnarray}
 \epsilon \tilde{h}^{ij} \equiv g^{ij} - \hat{g}^{ij}, \label{h_upper_indices2} \\
 \tilde{g}^{ij} \equiv \hat{g}^{ij} - \eta^{ij}, \label{g_bar_upper_indices2}
\end{eqnarray} 
from which one obtains the exact identity
\beq\label{g_upper_indices}
g^{ij} \ = \  \hat{g}^{ij} + \epsilon \tilde{h}^{ij}   \ = \ \eta^{ij} + \tilde{g}^{ij} + \epsilon \tilde{h}^{ij}.
\eeq
Now, $\epsilon\tilde{h}^{ij}$ and $\tilde{g}^{ij}$ are both of order $\epsilon$ because substituting \eqref{inverse perturbed metric} into \eqref{h_upper_indices2} gives 
\beq\label{tilde_h_order_epsilon} 
\epsilon\tilde{h}^{ij} = - \epsilon h^{ij} + O(\epsilon^2) = O(\epsilon),
\eeq
and similarly approximating $\hat{g}^{ij}$ in \eqref{g_bar_upper_indices2} by \eqref{g_bar upper indices} leads to 
\beq\nonumber \label{tilde_g_order_epsilon}
\tilde{g}^{ij}= - \bar{g}^{ij} + O(\epsilon^2) =O(\epsilon).
\eeq 
Regarding derivatives of $g^{ij}$, there is no reason to assume that $\hat{g}^{ij}_{\ \, ,l} = \tilde{g}^{ij}_{\ \, ,l}$ is $O(\epsilon)$, so we must keep all such terms in the linearized equations. However, $\epsilon \tilde{h}^{ij}_{\ \, ,l}=O(\epsilon)$, since the derivative of the inverse metric is given by 
$$
g^{ij}_{\ \, ,l} = - g^{i\sigma} g^{j\rho} g_{\sigma\rho,l} ,
$$ 
which implies by \eqref{h_upper_indices2} that
\begin{eqnarray}\label{dell_h_order_epsilon_eqn}
\epsilon\, \tilde{h}^{ij}_{\ \, ,l} 
&=& \partial_l \left( g^{ij} - \hat{g}^{ij} \right) \cr
&=& - g^{i\sigma} g_{\sigma\rho,l} g^{j\rho} + \hat{g}^{i\sigma} \hat{g}_{\sigma\rho,l} \hat{g}^{j\rho} \cr 
&=& \hat{g}^{i\sigma} \hat{g}^{j\rho} \left( \hat{g}_{\sigma\rho,l}  - {g}_{\sigma\rho,l}  \right) - \epsilon \left( \hat{g}^{i\sigma} \tilde{h}^{j\rho} + \tilde{h}^{i\sigma} \hat{g}^{j\rho} \right) g_{\sigma\rho,l} ,
\end{eqnarray}
and using now \eqref{tilde_h_order_epsilon} and
$$
\hat{g}_{\sigma\rho,l}  - {g}_{\sigma\rho,l} = \epsilon h_{\sigma\rho,l} = O(\epsilon),
$$ 
we finally obtain
\beq\label{dell_h_order_epsilon}
\epsilon \tilde{h}^{ij}_{\ \, ,l} = O(\epsilon).
\eeq
From this we conclude, that the inverse of $g_{ij}$ in the form \eqref{g_upper_indices} separates off the $O(1)$ term $\eta^{ij}$ from the $O(\epsilon)$ terms $\epsilon\tilde{h}^{ij}$ and $\tilde{g}^{ij}$, and \eqref{dell_h_order_epsilon_eqn} confirms that raising the indicies on $h$ keeps $h$ and derivatives of $h$ order $1$. 

Using \eqref{g_upper_indices}, we write the wave gauge condition in its equivalent form
\beq\label{wave_gauge_2}
\eta^{\sigma\rho} \tilde{\Gamma}_{i\, \sigma\rho} = - \left(\tilde{g}^{\sigma\rho} + \epsilon \tilde{h}^{\sigma\rho} \right) \tilde{\Gamma}_{i\, \sigma\rho}.
\eeq
Now, substituting \eqref{wave_gauge_2} into \eqref{Prop1_2derivative-term_eqn2a} we obtain
\beq\label{Prop1_2derivative-term_eqn3b} 
\eta^{ik} \tilde{\Gamma}_{i\, j[k,l]} 
= \frac12 \Box_\eta h_{jl} + \frac12 \partial_l \left( \big(\tilde{g}^{\sigma\rho} + \epsilon \tilde{h}^{\sigma\rho} \big) \tilde{\Gamma}_{j \, \sigma\rho} \right) + \frac12 \partial_j \left( \big(\tilde{g}^{\sigma\rho} + \epsilon \tilde{h}^{\sigma\rho} \big) \tilde{\Gamma}_{l \, \sigma\rho} \right),
\eeq
so that using \eqref{dell_h_order_epsilon}, $\hat{g}^{ij}_{\ \, ,l} = \tilde{g}^{ij}_{\ \, ,l}$ and $\epsilon \tilde{h}^{ij} = O(\epsilon) = \tilde{g}^{ij}$ finally leads to
\beq\label{Prop1_2derivative-term_eqn3} 
\epsilon \eta^{ik} \tilde{\Gamma}_{i\, j[k,l]}  =  \frac\epsilon2 \Box_\eta h_{jl} + \frac\epsilon2 \hat{g}^{\sigma\rho}_{\ \, ,l} \, \tilde{\Gamma}_{j \, \sigma\rho}  + \frac\epsilon2 \hat{g}^{\sigma\rho}_{\ \, ,j} \, \tilde{\Gamma}_{l \, \sigma\rho}  + O(\epsilon^2).
\eeq

To finish the proof, we combine the derivatives of $\hat{g}^{ij}$ in \eqref{Prop1_2derivative-term_eqn3} with $\{\cdot\}_\text{III}$ and substitute \eqref{g_bar upper indices} for $\hat{g}^{ik}$ which yields up to $O(\epsilon^2)$ errors that
\begin{eqnarray}\nonumber 
& &  \{\cdot\}_\text{III}  + \frac\epsilon2 \hat{g}^{\sigma\rho}_{\ \, ,l} \,\tilde{\Gamma}_{j \, \sigma\rho}  + \frac\epsilon2 \hat{g}^{\sigma\rho}_{\ \, ,j} \,\tilde{\Gamma}_{l \, \sigma\rho} \cr
&&=  \epsilon \,\left( \tilde{\Gamma}^k_{\ \sigma l} \bar{\Gamma}^\sigma_{\ jk} - \tilde{\Gamma}^k_{\ \sigma k} \bar{\Gamma}^\sigma_{\ jl} + \bar{\Gamma}^k_{\ \sigma l} \tilde{\Gamma}^\sigma_{\ jk} - \bar{\Gamma}^k_{\ \sigma k} \tilde{\Gamma}^\sigma_{\ jl} 
+ \frac12 \hat{g}^{\sigma\rho}_{\ \, ,l} \tilde{\Gamma}_{j \, \sigma\rho}  + \frac12 \hat{g}^{\sigma\rho}_{\ \, ,j} \tilde{\Gamma}_{l \, \sigma\rho} \right)   \cr
&&=\epsilon\, \left\{\delta^{\tau}_{l}\bar{\Gamma}^{\rho}_{j\sigma}
- \delta^{\tau}_{\sigma}\bar{\Gamma}^{\rho}_{jl}
+ \delta^{\rho}_{j}\bar{\Gamma}^{\tau}_{\sigma l}
-\delta^{\rho}_{j}\delta^{\tau}_{l}\bar{\Gamma}^{k}_{\sigma k}  
+ \frac12\left( \eta_{j\sigma} \hat{g}^{\rho\tau}_{\ \ ,l}   
+ \eta_{l\sigma} \hat{g}^{\rho\tau}_{\ \ ,j} \right) \right\}\tilde{\Gamma}^{\sigma}_{\rho\tau}, 
\end{eqnarray}
where we raise indices on $\tilde{\Gamma}$ and $\bar{\Gamma}$ with the Minkowski metric. Comparing the terms in the braces of the previous equation with the definition of the coefficients ${b}_{jl \sigma}^{\rho\tau}$ in \eqref{def_b} immediately yields
\beq\label{finaleqn_Prop1_c}
\{\cdot\}_\text{III}  + \frac\epsilon2 \hat{g}^{\sigma\rho}_{\ \, ,l} \tilde{\Gamma}_{j \, \sigma\rho}  + \frac\epsilon2 \hat{g}^{\sigma\rho}_{\ \, ,j} \tilde{\Gamma}_{l \, \sigma\rho} 
= \epsilon\, {b}_{jl \sigma}^{\rho\tau}\tilde{\Gamma}^\sigma_{\ \rho\tau}.
\eeq
Finally, combining \eqref{finaleqn_Prop1_c} with \eqref{Prop1_2derivative-term_eqn3}, we obtain \eqref{claim2}. This proves the Lemma and completes the proof of Theorem \ref{Prop1}. 
\QED

\subsection{Coriolis Accelerations for Gravitational Waves in Approximate Locally Inertial Frames }\label{sec: effect on high frequency waves}\label{coriolish}

Our goal here is to prove that in each approximate locally inertial frame there exists gravity waves $h_{ij}$ in the wave gauge which solve \eqref{EFE phys spacetime 4}, such that $\mathcal{C}_{jl}(h)$ is of order $O(1)$ in a neighborhood of $p$. In this case, $\mathcal{C}_{jl}(h)$ represents an acceleration to the gravity wave $h_{ij}$ which we identify as a non-removable GR Coriolis acceleration in analogy to the classical Coriolis force.

We see from (\ref{EFE phys spacetime 4}) that even though $\bar{g}$ is only Lipschitz continuous and the linearized Einstein equations contain second derivatives of $\bar{g}$, the delta functions cancel out in every approximate locally inertial frame because they only appear through the curvature tensor.    Due to cancellation, the curvature tensor does not distinguish an approximate locally inertial frame from an actual one.
This then begs the question as to whether it is possible for the first-order derivatives to similarly cancel in approximate locally inertial frames.  Asked differently, do the first-order derivatives of $\bar{g}$ distinguish approximate locally inertial frames from actual ones?  The purpose of this section is to answer this question in the affirmative by isolating a non-vanishing scattering effect in approximate locally inertial coordinate frames which distinguish them from actual locally inertial frames, and thereby complete the proof of Theorem \ref{Thmintro} of the Introduction. 

\begin{Thm}\label{thmclk} 
Under the assumptions of Theorem \ref{Prop1}, in each approximate locally inertial coordinate system at $p$, in the sense of Definition \ref{defapp}, there exists indices $j,l,\sigma,\rho,\tau$ with $\rho\neq\tau$, such that  
\beq\label{Cnonzero}
|b^{\rho\tau}_{jl\sigma}| \geq  \frac{1}{4}M .
\eeq
Moreover, within the class of initial data satisfying the wave gauge  \eqref{wave gauge non-linear},      there exist solutions $h$ of \eqref{EFE phys spacetime 4} and $l,j$ from \eqref{Cnonzero} such that 
\begin{eqnarray}\label{Cnonzero2}
|\mathcal{C}_{jl}(h)|&\geq & \frac{1}{8}M ,
\end{eqnarray}
in some open subset of ${U}$.
\end{Thm}

The theorem makes things simpler than one might expect because it applies whenever {\it any} derivative of  $\bar{g}$ fails to vanish.  

\Proof 
The well-posedness for \eqref{EFE phys spacetime 4} follows from the standard existence theory for linear hyperbolic PDE's of second-order \cite{Evans,Choquet}. It remains to prove \eqref{Cnonzero} and to show that there exists initial data for which the resulting solution satisfies \eqref{Cnonzero2}.  

We first prove \eqref{Cnonzero}. By \eqref{nonzeroderiv} of Definition \ref{defapp}, there exist indices $\alpha,\beta,\gamma$ for which $|\bar{g}_{\alpha\beta,\gamma}| > M$ in $\mathcal{U}$.  Now, since $\bar{g}_{\alpha\beta,\gamma}$ can be expressed as
$$
\bar{g}_{\alpha\beta,\gamma}=\bar{\Gamma}_{\alpha\beta\gamma}+\bar{\Gamma}_{\beta\gamma\alpha},
$$
we know that
$$
|\bar{\Gamma}_{\alpha\beta\gamma}+\bar{\Gamma}_{\beta\gamma\alpha}| 
=  |\bar{g}_{\alpha\beta,\gamma}|
> M ,
$$
yielding that there exists some set of indices, which for simplicity we again label as $\alpha,\beta,\gamma$, such that
\begin{eqnarray}\label{OneGammabig}
|\bar{\Gamma}^{\alpha}_{\beta\gamma}|\geq M/2.
\end{eqnarray}

We next verify that \eqref{OneGammabig} alone implies \eqref{Cnonzero}. To start, recall the definition of the coefficients $b_{jl\sigma}^{\rho\tau}$, \eqref{def_b}:
\begin{eqnarray}\nonumber 
b_{jl\sigma}^{\rho\tau}=\delta^{\tau}_{l}\bar{\Gamma}^{\rho}_{j\sigma}
-\delta^{\tau}_{\sigma}\bar{\Gamma}^{\rho}_{jl}
+\delta^{\rho}_{j}\bar{\Gamma}^{\tau}_{\sigma l}
-\delta^{\rho}_{j}\delta^{\tau}_{l}\bar{\Gamma}^{k}_{\sigma k}  +  
\frac12\left( \eta_{j\sigma} \hat{g}^{\rho\tau}_{\ \ ,l}   + \eta_{l\sigma} \hat{g}^{\rho\tau}_{\ \ ,j} \right).
\end{eqnarray}
To prove that \eqref{OneGammabig} implies \eqref{Cnonzero}, we compute $b_{jl\sigma}^{\rho\tau}$ under four different conditions on the indices $l,j,\sigma,\rho,\tau$.  For these explicit cases, we do {\it not} use the summation convention, but rather calculate $b_{jl\sigma}^{\rho\tau}$ for fixed indices $l,j,\sigma,\rho,\tau$ when explicit relations between them are assumed. 
\vspace{.2cm}  

\noindent Case I:  Assume $\sigma=\tau$, $\rho\neq\sigma$, $\sigma\notin\left\{j,l\right\}$, and $\rho\notin\left\{j,l\right\}$. Then by \eqref{def_b}
\begin{eqnarray}\label{caseI}
b_{jl\sigma}^{\rho\sigma}=0-\delta^{\sigma}_{\sigma}\bar{\Gamma}^{\rho}_{jl}+0-0 + \frac12 0 =\bar{\Gamma}^{\rho}_{jl}.
\end{eqnarray}

\noindent Case II:  Assume $\rho=l$, $\tau=\sigma$, $j\neq l$, and $\sigma\notin\left\{j,l\right\}$.  Then by \eqref{def_b}
\begin{eqnarray}\label{caseII}
b_{jl\sigma}^{l\sigma}=0-\delta^{\sigma}_{\sigma}\bar{\Gamma}^{l}_{jl}+0-0 + \frac12 0 =-\bar{\Gamma}^{l}_{jl}.
\end{eqnarray}
\vspace{.1cm} 
\noindent The constraints on the indices in Cases I and II together with the symmetry $\bar{\Gamma}^{\rho}_{lj}=\bar{\Gamma}^{\rho}_{jl}$ allow us to solve for any $\bar{\Gamma}^{\rho}_{jl}$ on the right hand side of either (\ref{caseI}) or (\ref{caseII}), except for the case when $\rho=j=l$.   To address this last possibility we require one additional case:
\vspace{.2cm}  

\noindent Case III:  Assume $\rho=l$, $\tau=\sigma$, $\sigma\neq l=j$.   Then by \eqref{def_b}
\begin{eqnarray}\label{caseIV}
b_{jl\sigma}^{l\sigma}=0-\delta^{\sigma}_{\sigma}\bar{\Gamma}^{l}_{jl}+ \delta^{l}_{j}\bar{\Gamma}^{\sigma}_{\sigma l}-0 + \frac12 0 = -\bar{\Gamma}^{l}_{ll} + \bar{\Gamma}^{\sigma}_{\sigma l}.
\end{eqnarray}
\vspace{.1cm}
Now assume $\bar{\Gamma}^{\alpha}_{\beta\gamma}\geq M/2$ for some values of $\alpha,\beta,\gamma$ such that not all of $\alpha,\beta,\gamma$ are equal, then, (since $\bar{\Gamma}$ is symmetric in the lower two indices), either $\alpha\notin\left\{\beta,\gamma\right\}$ or $\beta\neq\gamma,\alpha=\gamma$.  If the possibility $\alpha\notin\left\{\beta,\gamma\right\}$ holds, then we can apply (\ref{caseI}) of Case I to conclude
$$
|b_{\beta\gamma\sigma}^{\alpha\sigma}| = 
|\bar{\Gamma}^{\alpha}_{\beta \gamma}| 
> \frac12 {M},
$$
for some $\sigma\neq\alpha$, (again, we do not sum over $\sigma$ in the above equation).  If the possibility $\beta\neq\gamma$ and $\alpha=\gamma$ holds, then we can apply (\ref{caseII}) of Case II to conclude 
$$
|b_{\alpha\beta\sigma}^{\alpha\sigma}|
= |\bar{\Gamma}^{\alpha}_{\beta \alpha}|
> \frac12 M
$$
for $\sigma\notin\left\{\alpha,\beta\right\}$, and hence for some $\sigma\neq\alpha$.  This confirms \ref{Cnonzero} for every possibility except the case $\alpha=\beta=\gamma$.  So assume $\alpha=\beta=\gamma$.  Using this in (\ref{caseIV}) of Case III with $k\neq\alpha$, we conclude
\begin{eqnarray}\label{lastcase}
\bar{\Gamma}^{\alpha}_{\alpha \alpha}=\bar{\Gamma}^{k}_{k \alpha}-b_{\alpha\alpha k}^{\alpha k}.
\end{eqnarray}
Now, substituting \eqref{caseII} for some $\sigma \notin \{\alpha, k\}$ and $k\neq\alpha$, (that is, $b_{\alpha k \sigma}^{k\sigma}=-\bar{\Gamma}^{k}_{\alpha k}$), into \eqref{lastcase} yields
\beq\label{verylastcase}
\bar{\Gamma}^{\alpha}_{\alpha \alpha}=-b_{\alpha k \sigma}^{k\sigma} -b_{\alpha\alpha k}^{\alpha k},
\eeq 
which immediately implies
\begin{eqnarray}\label{lastcase3}
 | b_{\alpha k \sigma}^{k\sigma}  | + | b_{\alpha\alpha k}^{\alpha k} | 
\geq | b_{\alpha k \sigma}^{k\sigma} + b_{\alpha\alpha k}^{\alpha k} |
= |\bar{\Gamma}^{\alpha}_{\alpha \alpha}|
> \frac{M}{2} .
\end{eqnarray}
From this we conclude that either $ | b_{\alpha k \sigma}^{k\sigma}  | > {M}/{4} $ or $ | b_{\alpha\alpha k}^{\alpha k} | > {M}/{4} $. In Summary, we have proven that (\ref{Cnonzero}) holds for some coefficient $b_{jl\sigma}^{\rho\tau}$ such that $\rho\neq\tau$. 

To verify (\ref{Cnonzero2}), we need to prove that for some $j,l$, there is not sufficient cancellation among the expression for $\mathcal{C}_{jl}$ given in \eqref{def Coriolis term} to make $|\mathcal{C}_{jl}|< M/4$ for all $h$. To prove this, recall that by \eqref{Cnonzero}, there exists fixed indices $j,l,\sigma,\rho,\tau$ (with $\rho\neq\tau$) such that
\begin{eqnarray} \nonumber 
|b_{jl\sigma}^{\rho\tau}| > \frac{M}{4}.
\end{eqnarray}
Now, we first decompose $\mathcal{C}_{jl}$ as
\begin{eqnarray}\label{lastcase6}
\mathcal{C}_{jl}=b_{jl\sigma}^{\rho\tau}\left(\tilde{\Gamma}^{\sigma}_{\rho\tau}+Q(\tilde{\Gamma},h)\right),
\end{eqnarray}
where we define $Q$ to be the sum over all terms on the right hand side of \eqref{def Coriolis term}, which do not involve $\tilde{\Gamma}^{\sigma}_{\rho\tau}$, for the fixed indices $\sigma$, $\rho$, $\tau$, divided by the non-zero $b_{jl\sigma}^{\rho\tau}$ from \eqref{Cnonzero}. 

Now, to finish the proof, fix an arbitrary point $q \in U$. Then, by \eqref{lastcase6}, it remains only to show that there exist an $h$ such that, at the point $q$, $\tilde{\Gamma}^{\sigma}_{\rho\tau}$ satisfies
\beq\label{range for Gamma_tilde}
\frac{M}{8b} - Q \ < \ \tilde{\Gamma}^{\sigma}_{\rho\tau} \ < \ - \frac{M}{8b} - Q,
\eeq
for $b = | b_{jl\sigma}^{\rho\tau} |$.  Indeed, \eqref{range for Gamma_tilde} is the sought after estimate \eqref{Cnonzero2} at the point $q \in {U}$, which by continuity would hold on some neighborhood of $q$. Therefore, it remains only to show that there exists enough freedom to choose $h$ and derivatives of $h$ at the single point $q$ to make $\tilde{\Gamma}^{\sigma}_{\rho\tau}$ satisfy \eqref{range for Gamma_tilde} at  $q$.  For this final step, the only restriction among all $\tilde{\Gamma}^l_{ij}$ are the Einstein constraint equations and the wave gauge condition \eqref{wave gauge non-linear}, because $\tilde{\Gamma}^l_{ij}$ at $q$ can be considered to be initial data from which $h$ is determined. 
              
The rigorous analysis of the constraint equations is beyond the scope of this paper, but all we are assuming here is that assigning some value for the $\tilde{\Gamma}^{\sigma}_{\rho\tau}$, at the single point $q$, is consistent with the well-posedness of the constraint equations. 

Finally, by \cite{Choquet}, the solution $h_{ij}$ is determined by initial data $h_{\alpha\beta}$ and $h_{\alpha\beta,0}$ for $\alpha,\beta \in \{1,2,3\}$ on $\Sigma=\{x^0=0\}$, leaving the freedom to arbitrarily assign $h_{i0}$ and $h_{i0,0}$ for $i\in \{0,..,3\}$. These are determined by the wave gauge conditions \eqref{wave gauge non-linear} in its equivalent form
\beq\nonumber
\epsilon\, g^{k0}\left(h_{ik,0} + h_{i0,k} - h_{k0,i} \right) = \sum_{\alpha,\beta=1}^3 g^{\alpha\beta}\tilde{\Gamma}_{i\, \alpha\beta},
\eeq 
where the right hand side only depends on spatial components of $h$. Thus, the wave gauge conditions \eqref{wave gauge non-linear} can be satisfied by choosing $h_{i0,0}$ accordingly (c.f. p. 164, chapter 7, in \cite{Choquet}), which then leaves $h_{\alpha\beta}$ and $h_{\alpha\beta,0}$ for $\alpha,\beta \in \{1,2,3\}$ free to assign as initial data and this freedom clearly suffices to arrange for \eqref{Cnonzero2} to hold. In summary, we have shown that one can choose the value of $\tilde{\Gamma}^\sigma_{\rho\tau}$ at $q$ to satisfy \eqref{range for Gamma_tilde} at the start and then it is consistent to solve the wave gauge and the constraint equations simultaneously with that value. This completes the proof of the Theorem.  
\QED

\section{Conclusion}\label{conclusion}

The point of departure for this paper is the authors' recent work in \cite{ReintjesTemple2,Reintjes} which gives the first proof that spacetime is locally inertial at a point of shock wave interaction. The method in \cite{ReintjesTemple2,Reintjes} is tailored to the simplest case of shock wave interaction, to the case of interaction between two shock waves from different characteristic families in spherical symmetry, and it is an open problem whether regularity singularities exist for more complicated shock wave interactions. We know of no physical principle that can rule out regularity singularities ahead of time, and only a mathematical proof can resolve the problem as to whether regularity singularities exist.    In this paper we clarify and motivate the open problem by investigating the physical implications of regularity singularities should they in fact exist.  To clarify the issue, we make the distinction between the smooth atlas of $C^{2,1}$ coordinate transformations and the larger atlas of $C^{1,1}$ transformations.   We then prove that, restricting to the $C^{2,1}$ atlas for a metric that is Lipschitz continuous at a point of shock wave interaction, the closest one can get to a locally inertial coordinate system is one which is {\it approximate locally inertial} in a sense we clarify.   We then linearize the Einstein equations in an approximate locally inertial coordinate system, and identify and characterize the Coriolis type terms which we prove will only vanish in a true locally inertial coordinate system, {\it should one exist}.     The open problem of regularity singularities, then, is the problem as to whether the approximate locally inertial coordinate systems can be improved to locally inertial coordinate systems within the larger $C^{1,1}$ atlas.   If locally inertial coordinates do not exist within the $C^{1,1}$ atlas, then the scattering of gravitational radiation by a regularity singularity would produce quantifiable physical effects analogous to non-removable Coriolis type forces, and these Coriolis effects are characterized in this paper.

\providecommand{\bysame}{\leavevmode\hbox to3em{\hrulefill}\thinspace}
\providecommand{\MR}{\relax\ifhmode\unskip\space\fi MR }
\providecommand{\MRhref}[2]{%
  \href{http://www.ams.org/mathscinet-getitem?mr=#1}{#2}
}
\providecommand{\href}[2]{#2}

\end{document}